# A Bayesian Network-Driven Zero Trust Model for Cyber Risk Quantification in Small-Medium Businesses

[1,2]*Ahmed M. Abdelmagid, [3]Barry C. Ezell, [4]Michael McShane

[1]Engineering Management and Systems Engineering Department, Old Dominion University, Norfolk, Virginia 23508, USA

[2]Production Engineering Department, Alexandria University, Alexandria, Egypt.

[3]Virginia Modelling, Simulation, and Analysis Simulation Center (VMASC), Suffolk, Virginia 23435, USA.

[4]Department of Finance, Strome College of Business, Old Dominion University, Norfolk, Virginia 23508, USA.

## Abstract

Small-Medium Businesses (SMBs) are essential to global economies yet remain highly vulnerable to cyberattacks due to limited budgets, inadequate cybersecurity expertise, and underestimation of cyber risks. Their increasing reliance on digital infrastructures has expanded their attack surfaces, exposing them to sophisticated and evolving threats. Consequently, implementing proactive, adaptive security measures has become imperative. This research investigates the effectiveness of Zero Trust Architecture (ZTA) as a sustainable cybersecurity solution tailored to SMBs. While ZTA adoption has been examined broadly, the specific financial, organizational, and capability constraints of SMBs remain underexplored. This study develops an integrated predictive model to assess both the feasibility and risk-mitigation potential of ZTA implementation. The model consists of two interlinked Bayesian network sub-models: the first evaluates the probability of successful ZTA adoption considering implied barriers, and the second tests the effectiveness of ZTA in responding to prevalent cyberattacks. The integrated model predicts the risk level in the presence of ZTA and quantifies the uncertainty of the extent to which ZTA can enhance SMBs' cyber resilience, contributing novel insights for practitioners and stakeholders seeking to enhance compliance with policies, risk, and governance activities in SMBs.

## 1. Introduction

The global advancement in Information and Communication Technology (ICT) has transformed business operations, compelling organizations to adopt emerging digital solutions such as the Internet of Things (IoT) and cloud computing, particularly after the COVID-19 pandemic [1]. With over 2.3 billion people relying on online platforms for work, education, commerce, and daily transactions, digital connectivity has become fundamental to business continuity and competitiveness [2]. SMBs, in particular, have embraced digital transformation to enhance their presence in the global e-commerce environment. However, these technological gains come with heightened cybersecurity risks, as SMBs often operate with limited financial and human resources, leaving their networks highly exposed to malicious intrusions [3].

Cyberattacks on SMBs frequently result in severe financial and reputational damage, data exposure, and regulatory penalties. noncompliance with frameworks such as the General Data Protection Regulation (GDPR) in the European Union and the Health Insurance Portability and Accountability Act (HIPAA) in the United States can cost millions in fines, compounding the financial strain on smaller enterprises [4]. Given the pivotal role SMBs play in the global economy, constituting virtually all businesses in Australia, the United Kingdom, and the United States, and employing nearly half the workforce in these regions [5-7], the ripple effects of cyber incidents can extend to critical sectors including healthcare, energy, and finance [8].

Definitions of SMBs vary across jurisdictions and industries. For example, the Australian Bureau of Statistics designates businesses with 5–199 employees as SMBs [5], while the U.S. Small Business Administration (SBA) recognizes those with fewer than 500 employees [9], with thresholds adjusting by industry category under the North American Industry Classification System (NAICS). This research work considers the threshold of 1,500 employees as the maximum headcount within SMBs. Thus, the SBA's classification is employed to frame the dataset available for this research within the U.S. context.

Cybercrime has escalated dramatically over the past decade. Between 2014 and 2024, the proportion of organizations experiencing six or more cyberattacks nearly doubled, while global cybercrime costs surged from $3 trillion in 2015 to $9.5 trillion in 2024, with projections exceeding $10 trillion by 2025 [10]. The 2024 IBM Data Breach Investigation Report found that average breach costs reached $4.88 million, marking a record high [11]. Despite these alarming statistics, many SMBs continue to lack structured cybersecurity frameworks, adequate risk management processes, and trained personnel to respond effectively to evolving cyber threats [12]. Limited budgets and fragmented security strategies further hinder their ability to identify, protect, detect, and recover from attacks [13]. Consequently, developing robust, scalable, and economically viable cybersecurity models tailored to SMBs is essential for ensuring their resilience and sustaining economic stability in an increasingly digitalized world [14].

Despite the growing recognition of Zero Trust Architecture (ZTA) as a holistic cybersecurity paradigm, its adoption remains hindered by the absence of a standardized framework guiding organizations in its implementation [15]. Security practitioners and decision-makers lack structured methodologies and benchmarks for selecting appropriate Zero Trust measures, often relying on subjective judgment rather than evidence-based strategies [16]. This gap impedes leadership within organizations, particularly in SMBs, from effectively transforming conceptual security goals into actionable decisions [17]. Although various government directives and strategic initiatives advocate for ZTA adoption, practical guidance on its deployment across diverse operational contexts, such as those of SMBs, is still lacking.

ZTA is a cybersecurity paradigm grounded in the principle of "never trust, always verify," under which no entity is inherently trusted, and every access request to network resources must be continuously authenticated and authorized. ZTA, originally conceptualized by Kindervag [18, 19], emphasizes strong authentication, strict authorization, and the minimization of implicit trust, while still preserving resource availability and minimizing authentication latency. Access decisions are enforced through granular policies at the policy decision and enforcement points, where each subject's request to access an organizational resource is evaluated against least-privilege rules before approval or denial [20]. As shown in Figure 1, the architecture is organized around several pillars, each supported by multiple security tools that must be implemented to achieve effective operation. However, migrating to this model demands substantial financial investment and technical expertise, capabilities that are often limited in SMBs, and is further complicated by the absence of a standardized migration strategy tailored to SMB-specific constraints and characteristics.

ZTA could be employed to combat cyberattacks executed through different tactics, techniques, and procedures in addition to helping business organizations, federal agencies, and associations to mitigate cyber threats related to various industry sectors. For example, Abdelmagid, Javadnejad [21] proposed a study in which a new integrated framework that utilized Security Engineering Risk Analysis (SERA) and MITRE ATT&CK frameworks to aid maritime stakeholders in conducting cyber risk assessments and adopting best cybersecurity practices. The authors indicated that the migration to ZTA will confer an adequate level of cyber hygiene. Moreover, Javadnejad, Abdelmagid [22], who proposed an exploratory analysis through the same dataset used in this study, focusing on ransomware and malware attacks that targeted multiple industry sectors, referred to ZTA as a potential enabler in reducing the negative consequences of such attacks.

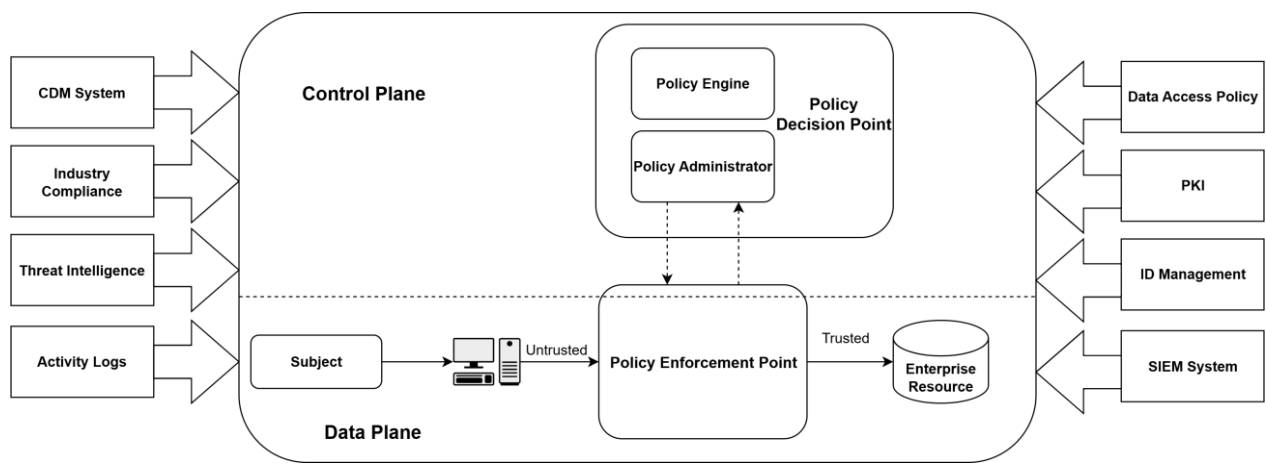

Figure 1 Zero Trust Architecture notion [20].

The core research problem addressed in this study lies in understanding and quantifying the factors influencing successful ZTA implementation within SMBs, considering their structural, financial, and resource constraints. SMBs face distinct challenges in migrating from legacy systems due to their limited budgets, workforce capabilities, and technological maturity. Yet, existing studies have not adequately examined how these constraints affect implementation success or how ZTA can mitigate cyber risk in this business segment. To fill this gap, this research investigates the interaction between organizational and financial barriers, relevant ZTA pillars, and the resulting effectiveness of Zero Trust measures in reducing risk magnitude through simulation-based analyses of real-world cyberattack data. The proposed study attempts to respond to the following research questions:

***RQ1: What is the impact of the financial and organizational barriers of ZTA on its adoption success?***

***RQ2: What are the relevant ZT security pillars and the associated security measures that could enhance SMBs' cybersecurity against certain cyber risks?***

***RQ3: How effective is the adoption of ZT security tools in the context of SMBs in reducing the magnitude of risk?***

***RQ4: How do various ZT controls affect the likelihood of data breach incidents in SMBs adopting ZTA?***

*RQ5: Which attack vectors pose the highest probability of a data breach incident, leading to the highest overall cyber risk level?*

This study makes significant theoretical, methodological, and practical contributions to the cybersecurity field by addressing the conspicuous gap in existing research on implementing ZTA within the SMBs' ecosystem. Theoretically, it integrates risk management principles into a Zero Trust framework to advance a risk-informed decision-making approach that supports governance, compliance, and proactive cyber risk strategies for SMBs. It further contributes by operationalizing the three fundamental pillars of cybersecurity, including People, Process, and Technology [23] within an integrated model that captures insider threat behavior, organizational and financial barriers, and technological enablers of ZTA, which is underrepresented in the literature [24]. Methodologically, it pioneers the use of Bayesian Network modeling to simulate cyberattack scenarios and quantify the uncertainty inherent in cybersecurity decisions, thereby predicting the effectiveness of ZTA in reducing cyber risk and identifying the key drivers of successful adoption. Practically, the model provides decision-makers with an evidence-based tool to evaluate ZTA's impact on their security posture, balance economic constraints, and select effective security controls. This research offers a comprehensive, data-driven framework to help SMBs transition toward Zero Trust principles to strengthen digital resilience against evolving cyber threats by bridging theoretical gaps and delivering actionable guidance.

The remainder of this paper is organized as follows: Section 2 illustrates the previous research studies related to the problem context and concludes the research gap, while Section 3 discusses the proposed model theoretical framework and the parameters considered. Section 4 introduces the modelling approach and simulation model. Section 5 demonstrates the obtained results, while Section 6 proposes the key research takeaways, conclusions, and future work.

## 2. Review of Literature

Cyber incidents affect organizations of all sizes, yet SMBs remain particularly vulnerable due to their weaker cybersecurity posture and limited resources. Unlike large enterprises that can allocate substantial budgets, hire skilled IT professionals, and rapidly respond to threats [25, 26], SMBs often lack the financial capacity and technical expertise necessary to implement effective cybersecurity measures. This resource deficit not only increases their exposure to attacks but also amplifies the financial consequences of successful breaches [27]. Hayes [28] claimed that cybercriminals frequently target SMBs under the assumption that their systems are easier to compromise, a claim supported by reports showing that smaller firms continue to lag in cybersecurity investment. According to Verizon's 2022 report, SMBs are often viewed as "low-hanging fruit" for attackers, facing common threats such as ransomware, phishing, and data theft incidents severe enough to drive many businesses out of operation [29]. Given these risks, understanding the barriers that prevent SMBs from strengthening their cybersecurity defenses is crucial, as illustrated in the next subsection.

### 2.1 Cybersecurity Challenges for SMBs

The literature identifies several impediments that hinder SMBs from adopting robust cybersecurity frameworks. These challenges are often categorized as financial, technical, organizational, human, and legal [4, 30]. Limited financial capacity restricts investment in cybersecurity tools or specialized personnel, while knowledge gaps exacerbated by the age or IT inexperience of business owners further weaken defensive capabilities [31].

### 2.1.1 Cyber Threats Underrating/Overconfidence

SMBs frequently underestimate the severity of cyber threats, assuming their data lacks sufficient value to attract attackers. This complacency leads many to neglect necessary precautions, underfund security initiatives, and overlook compliance with standards such as the Payment Card Industry Data Security Standard (PCI DSS) [32]. Surveys show that only a small proportion of SMBs allocate budgetary resources to cybersecurity or comply with essential data protection regulations [33], reflecting a widespread misjudgment of cyber risk consequences.

### 2.1.2 Lack of Awareness and Expertise

Weak cybersecurity culture stemming from poor risk perception, limited technical knowledge, and uninformed behavior further undermines SMBs' resilience [34]. Many employees and managers struggle to recognize cyber risks or assess third-party security services effectively [35]. The lack of skilled IT specialists also prevents the adoption of advanced monitoring systems such as Security Information and Event Management (SIEM), intensifying SMBs' operational exposure [26].

### 2.1.3 Limited Human and Financial Resources

Financial and staffing constraints remain critical barriers [36, 37]. SMBs face the same threats as large enterprises but with a fraction of the resources. High implementation costs often discourage investment in comprehensive defense systems, forcing SMBs to find cost-efficient, multi-functional cybersecurity solutions that balance affordability with reasonable protection [38, 39].

### 2.1.4 Speed Evolution of Digital Technology

The rapid evolution of digital technologies compounds these difficulties. As information and communication technology advances, cyberattacks become increasingly sophisticated, outpacing SMBs' capacity to update defenses [40]. The inability to keep pace with emerging threats further limits effective risk management, exposing SMBs to cascading vulnerabilities within modern digital ecosystems.

## 2.2 ZTA as a Risk Countermeasure for SMBs

### 2.2.1 ZT Pros SMBs

Cybersecurity resilience and profitability have emerged as strategic imperatives for modern enterprises [41]. The ZT security framework serves as a critical enabler in achieving these goals by strengthening protection mechanisms and optimizing operational efficiency. Technically, ZT mitigates lateral movement threats often associated with Virtual Private Network (VPN) vulnerabilities, thereby reducing the overall attack surface. Economically, it delivers measurable cost savings by minimizing dependence on VPNs, consolidating software licenses, and leveraging scalable clustering technologies. These attributes collectively enhance both network security and the return on investment. For SMBs, frequent onboarding of employees, contractors, and vendors introduces additional access points, increasing exposure to cyber threats. Contrary to the misconception that ZT hinders business agility [42], the model safeguards digital infrastructure, protects intellectual property, and reduces downtime from security incidents, ultimately supporting sustained business growth.

Kudrati and Pillai [42] identify several key ZT controls suitable for SMBs, including Multi-Factor Authentication (MFA) to prevent privilege escalation, password vaults for securing shared credentials, and policies of least privilege and secure remote access to restrict unnecessary data exposure. Continuous monitoring and auditing help identify malicious behavior and ensure compliance with frameworks such as HIPAA, FISMA, and NIST [29], while Privileged Access Management protects critical administrative accounts. Collectively, these tools enable a "never trust, always verify" environment, reinforcing accountability across all network entities.

The adoption of ZT can reduce SMB dependence on external IT providers, accelerate incident response, and enhance regulatory compliance. For small businesses within regulated sectors, such as defense contractors required to meet the Cybersecurity Maturity Model Certification (CMMC), ZT offers a practical pathway toward secure cloud-based compliance and operational integrity [35]. Notably, while global ZT adoption continues to grow, smaller enterprises remain less likely than large corporations to execute full-scale migration plans, underscoring the need for tailored models that address their unique financial and technical constraints [43].

2.2.2 ZT Cons for SMBs

Organizations face multiple obstacles when transitioning to the ZT security framework, largely due to the extensive structural, financial, and operational changes required. Implementing ZT often entails replacing legacy systems, restructuring IT networks, and enforcing complex access policies for diverse user groups, which can be resource-intensive and time-consuming. The associated costs of new technologies, specialized training, and workflow adaptation frequently discourage adoption. Moreover, uncertainty regarding the Return On Investment (ROI) and the difficulty of assessing the financial benefits further complicates decision-making, particularly for SMBs with limited budgets and cybersecurity expertise. These challenges can inadvertently heighten organizational exposure to advanced cyber threats despite ZT's potential advantages [44, 45].

Conceptual ambiguities surrounding ZT also hinder its widespread implementation. Although frameworks such as NIST SP 800-207 provide general guidelines, critical ambiguities persist regarding levels of trust, component interaction, and integration practices. Michael, Dinolt [15] argue that ZT's underlying principles can be inconsistently interpreted, leading to subjective implementations and a lack of standardization across organizations. They noted that continuous maintenance, configuration updates, and monitoring add operational burdens that contradict the "economy of mechanism" design principle advocating simplicity [15].

Despite these limitations, ZT remains a vital and adaptive approach for combating evolving cyber threats. Recent analyses underscore its growing importance for U.S. SMBs, where ZT adoption is increasingly tied to national security and economic resilience [46]. To mitigate resource constraints, researchers recommend that governments encourage adoption through measures such as cybersecurity investment tax credits, while vendors focus on providing cost-effective, scalable ZT solutions for smaller enterprises. A detailed discussion on the ZT advantages and disadvantages can be found in [47].

2.2.3 ZTA Implementation Risks

Implementing ZTA represents a fundamental shift in cybersecurity, emphasizing least-privilege access and rigorous verification. However, this transition introduces a range of implementation risks, particularly for SMBs that operate under financial and resource constraints. Among these, financial and organizational risks are the most critical and recurrent barriers to effective ZT adoption.

2.2.3.1 Financial Risks

ZTA implementation frequently demands considerable capital investment due to the need to replace or upgrade legacy IT and operational technology (OT) systems with newer, ZT-compatible solutions [48]. These expenditures extend beyond hardware and software acquisition, encompassing continuous administrative upkeep, policy revisions, and visibility monitoring, often requiring additional personnel and inflating operational overhead [49]. Moreover, maintenance and technical support costs must be sustained to ensure long-term reliability. A major challenge arises from inaccurate budget estimations and uncertainty regarding the ROI.

The inherent lack of standardized cost frameworks and limited understanding of ZT's core concepts complicate financial forecasting, making it difficult for organizations to quantify potential cost savings or risk reduction benefits [50-53]. For SMBs, constrained budgets and limited technical expertise magnify these issues, forcing decision-makers to balance cyber protection needs against competing financial priorities. Unclear ROI projections often lead to hesitation or partial implementation, thus undermining the intended security benefits of the ZT model.

2.2.3.2 Organizational Risks

Beyond financial barriers, organizational factors also hinder successful ZTA deployment. Integrating ZTA into environments built around outdated systems frequently proves challenging, as legacy infrastructures often lack the required authentication and access controls [48]. Replacing or retrofitting these systems is costly, time-consuming, and operationally disruptive, with studies showing that nearly half of ZT adopters identify legacy integration as the most substantial threat to implementation success [54]. Furthermore, resistance to change within organizations poses a persistent obstacle. Employees and managers may perceive new security measures, such as multifactor authentication and continuous monitoring, as intrusive or workflow-restrictive [52]. Without a robust change management plan and user engagement, this cultural resistance can jeopardize adoption efforts. In addition, ZT integration across multiple bundled security solutions often introduces configuration complexities that demand specialized expertise [48]. Decision paralysis, resulting from the perceived difficulty and novelty of ZT, further delays strategic progress, especially among smaller firms lacking cyber risk comprehension [55-57].

2.2.3.3 Other Risks

While financial and organizational factors dominate, additional risk dimensions merit brief consideration. Technically, ZT deployment may expose vulnerabilities related to credential theft, network visibility gaps, and overreliance on proprietary tools [20]. Supply chain risks also emerge when SMBs act as vendors within larger ecosystems, where insufficient alignment with ZT standards can expose interconnected networks to cascading breaches [58, 59]. These external dependencies amplify vulnerability, particularly across complex multi-tier supplier arrangements [60-62]. ZTA offers transformative security benefits, but its adoption within SMBs is frequently constrained by financial burdens, unclear ROI, legacy system incompatibility, and internal resistance to change. Overcoming these challenges requires adopting phased migration strategies, ensuring vendor interoperability, and developing standardized cost assessment frameworks that make ZT both feasible and sustainable for resource-limited organizations.

Readers are referred to the research article introduced by Abdelmagid [59], who provided a thorough discussion of ZT implementation risks from the perspective of SMBs and developed a taxonomy that classifies all risks.

## 2.3 Research Gap and Summary

A review of academic and industry literature related to ZTA implementation in SMBs' context reveals increasing attention to ZTA as a modern cybersecurity solution. Although its advantages are widely discussed, the analysis shows that research remains heavily skewed toward general observations about ZTA's benefits, while its risks and implementation challenges receive limited emphasis. Out of 49 reviewed sources, only 15 examined ZTA in the context of SMBs, and the majority focused on highlighting advantages rather than barriers or risk mitigation. Financial and organizational obstacles are occasionally mentioned, and literature lacks investigation into technical, operational, and supply chain risks relevant to SMBs.

Existing research efforts, including those by Luckett [63], Kudrati and Pillai [42], Adahman, Malik [64], and Aurelien [61], provide only partial perspectives. While they recognize ZTA's strategic role in strengthening SMB cybersecurity, these works stop short of offering comprehensive frameworks that address financial and organizational barriers or measure ZTA's effectiveness in reducing cyber risks. Specifically, Luckett [63] focuses on guidelines and roadmaps for adoption but overlooks the mitigation of resource-based risks, while Kudrati and Pillai [42] identify certain ZT-related risks without examining their broader implications. Similarly, Adahman, Malik [64] emphasize cost analysis but neglect integration challenges particular to SMBs, and Aurelien [65] highlights ZTA's potential without assessing its disadvantages.

The core research gap identified lies in the absence of a holistic and quantitative approach to studying ZTA implementation within SMBs. Existing literature does not adequately explain how SMBs can adopt ZTA while managing inherent financial, organizational, and residual risks. Furthermore, no established model currently quantifies the likelihood of successful ZT adoption or its impact on mitigating cyber threats within SMB environments. This research addresses that gap by developing a Bayesian-based predictive model to evaluate the feasibility and risk mitigation potential of ZTA for SMBs, thereby contributing both theoretical and practical insights into the advancement of Zero Trust–driven cybersecurity for resource-constrained organizations (e.g., SMBs).

## 3. The Proposed Model

### 3.1 The Modeling Approach and Rationale

Bayesian Network Modeling is particularly well-suited to address the objectives of this study, since, unlike traditional statistical techniques that rely heavily on large and complete datasets, BNs enable the integration of limited empirical evidence with expert knowledge, which is an essential capability in cybersecurity, where historical data are often scarce and evolving threats limit the availability of consistent observations. Through their probabilistic framework, BNs provide a rigorous mechanism for constructing models that reflect both known evidence and expert judgment, resulting in a coherent representation of uncertainty within complex decision environments.

A distinct advantage of BNs lies in their ability to model causal relationships among interdependent factors, allowing researchers to examine how various elements, such as organizational characteristics, financial constraints, and the deployment of ZT technologies, influence cybersecurity outcomes. This causal representation supports scenario analysis and inference under uncertainty, enabling predictions of how specific interventions, such as budget adjustments or capability enhancements, may affect ZT implementation success. Furthermore, BNs are inherently dynamic, updating prior beliefs as new data becomes available, thereby ensuring adaptability and continuous refinement of the model. Given these strengths, BNs provide a superior approach for analyzing ZT adoption in SMBs, where the interplay of limited data, evolving risks, and organizational diversity creates an environment of high uncertainty. Consequently, this study employs Bayesian Network modeling as its analytical foundation.

**3.2 Overview of BN Modeling Approach**

A Bayesian Network (BN) is a graphical model that represents probabilistic dependencies among a set of variables [66]. It employs conditional probabilities to establish relationships between variables and applies Bayes' theorem and associated Bayesian learning algorithms to compute the posterior probabilities of outcome states [67]. As a subset of Artificial Intelligence (AI) techniques, BNs can learn probabilistic associations and interdependencies among variables using machine learning algorithms, thereby enabling reasoning and decision-making under uncertainty, a defining characteristic of AI systems. Through Bayesian inference, BNs update prior beliefs as new evidence emerges, making them particularly valuable for predictive analysis and diagnostic applications in dynamic decision contexts [68, 69].

BNs can be categorized into dynamic and static types [70]. Dynamic BNs capture time-dependent variations and evolving relationships between system variables, while static BNs analyze risk states and identify critical components without modeling temporal interactions. Structurally, a static BN comprises two components: a qualitative part, represented by a Directed Acyclic Graph (DAG) containing nodes (random variables) and edges (conditional dependencies), and a quantitative part consisting of conditional probability distributions that define the strength of these relationships [24]. BNs can be constructed either from the ground up using existing theoretical or empirical knowledge to define variables and causal links [71, 72] or derived from established methods such as Fault Trees and Bow Tie models. In some cases, data mining algorithms may automate the discovery of dependencies between variables, while expert judgment or prior knowledge serves to refine and validate the model structure [70]. The knowledge elicited from subject matter experts [70] or the domain knowledge of the author [24] can be employed to come up with the relevant variables and their causal relationship to form an informative BN structure that represents the studied system.

In cybersecurity risk assessment, BNs are typically built manually, as automatic learning techniques require extensive datasets that are often unavailable in this field [70]. The scarcity of comprehensive data on cyber incidents and breaches presents a major limitation for training data-driven models. Accordingly, BNs are particularly advantageous because they can integrate diverse information sources, empirical data, expert opinions, and theoretical constructs, to model uncertainty and infer potential cyber risk scenarios even when historical data is limited [24].

The British mathematician Thomas Bayes introduced the mathematical concept of BN in the 17th century [73]. Hossain, Nur [74] elaborated the mathematical concept of the Bayesian network as follows:

The probability of two events, $A$ and $B$, occurring together is the product of the probability of $A$ and the probability of $B$ given $A$, as illustrated in Equation 1.

$$P(A, B) = P(A) \times P(B|A) \qquad \text{(Equation 1)}$$

where $P(A, B)$ is the probability of both events occurring simultaneously. Equation 1 can be altered by the law of symmetry and rewritten as in Equation 2.

$$P(A|B) = \frac{P(B|A) \times P(A)}{P(B)} \qquad \text{(Equation 2)}$$

Where $P(A)$ is the prior probability of event $A$, $P(B)$ is the prior probability that evidence $B$ is true, $P(A|B)$ is the posterior probability of $A$ given evidence $B$, and $P(B|A)$ is the probability of evidence $B$ if hypothesis $A$ is true.

Assume a BN structure includes $n$ variables, $A_1, A_2, A_3, \ldots, A_n$. The general form to determine the joint probability distribution of the variables can be deduced as in Equation 3. Equation 3 can then be streamlined to Equation 4.

$$P(A_1, A_2, A_3, \ldots, A_n) = P(A_1| A_2, A_3, \ldots, A_n) P(A_2| A_3, \ldots, A_n) \ldots P(A_{n-1}| P_n) P(A_n) \quad \text{(Equation 3)}$$

$$P(A_1, A_2, A_3, \ldots, A_n) = \prod_{i=1}^{n} P(A_i| A_{i+1}, A_{i+2}, \ldots, A_n)$$
$$= \prod_{i=1}^{n} P(A_i| Parents(A_i)) \quad \text{(Equation 4)}$$

Figure 2 illustrates a Bayesian Network (BN) composed of four variables, $S = \{A_1, A_2, A_3, A_4\}$, where a series of arcs denotes the causal dependencies among them. An arc directed from one variable $A_i$ to $A_j$ signifies that the value of the latter depends on the former, thereby establishing an interdependent relationship between the two variables. Every BN structure is composed of three categories of nodes: (1) nodes without any parent nodes, referred to as root nodes; (2) nodes that lack child nodes, known as leaf nodes; and (3) nodes that possess both parent and child nodes, defined as intermediate nodes. Paret nodes are the origin points from which arcs extend to connect with their corresponding child nodes, while child nodes are those that receive incoming arcs from parent nodes [70]. In Figure 2, the nodes labeled as $A_1$ and $A_3$ have no parent nodes and are therefore identified as root nodes, whereas $A_4$ is considered a child node since it lacks any leaf nodes, and $A_2$ serves as an intermediate node. The general formula representing the complete joint probability distribution for the BN is expressed in Equation 5.

$$P(A_1, A_2, A_3, A_4) = P(A_1)P(A_3)P(A_2| A_1)P(A_4| A_2, A_3)$$
$$= \sum P(A_j|Parents(A_i)) \quad \text{(Equation 5)}$$

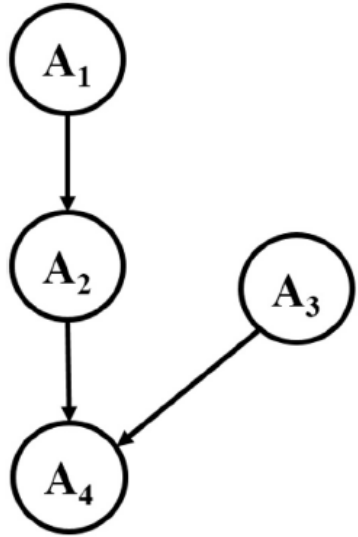

Figure 2 An example of a BN that contains four variables [74].

The versatility and effectiveness of Bayesian Network (BN) modeling have been demonstrated across a wide range of practical applications and critical infrastructures. BNs have been used extensively for diagnostic and predictive purposes in fields where data limitations complicate traditional statistical modeling, including medical diagnostics [75], system failure analysis [75], water resource management [76], environmental monitoring [77], and agriculture [78]. In safety risk management, several studies have leveraged BNs to predict hazards and optimize preventive strategies. For example, Xin, Khan [79] employed BNs to identify chemical hazards through case studies consistent with reported incidents, while Mohammadfam, Ghasemi [80] used BNs to model worker safety behaviors in power plant construction, highlighting their ability to evaluate intervention measures. Similarly, Zhou, Li [81] applied BN-based risk modeling to assess diaphragm wall deflection during excavations, showing BNs' superiority over fault tree methods for real-time risk analysis. Other implementations include the evaluation of port resilience and interdependencies within supply chains [82], as well as modeling risks in oil and gas operations to enhance supply chain robustness [83].

Within the domain of cyber risk management, BN models have been extensively explored to predict and diagnose security vulnerabilities, classify threats, and evaluate risk mitigation strategies. Chockalingam, Pieters [24] categorized BN-based cyber risk models into diagnostic and predictive types. Predictive models aim to forecast potential causes and their effects on cybersecurity systems. For instance, Poolsappasit, Dewri [84] quantified the probability of network compromise at multiple layers, while Frigault and Wang [85] estimated the likelihood of successful attacks via vulnerability exploitation. In critical infrastructure security, Shin, Son [86] introduced a BN model for assessing nuclear reactor protection systems, identifying the most impactful cyberattacks and corresponding mitigation actions. Their subsequent research integrated BN outputs with event tree analysis to compute system failure probabilities and model the relationship between attacks and defensive responses [86].

Diagnostic BN models, though less prevalent, have also been applied to automate vulnerability classification and evaluate insider-driven data breaches [87, 88]. Collectively, these studies underscore the capability of BN modeling to quantify uncertainty, analyze causal dependencies, and enhance decision-making in complex cyber risk scenarios, establishing it as a powerful framework for cybersecurity risk assessment and management. Table 1 summarizes the applications found in the literature that utilize the BN modeling approach.

Table 1 A summary of the BNs' applications in the literature.

| Reference | Theme | Application Domain |
|---|---|---|
| **Hossain, Jaradat [184]**, **Shan, Liu [185]** | | Oil and Gas |
| **Xin, Khan [179]**, **Weber, Medina-Oliva [172]**, **Zhou, Li [181]**, **Mohammadfam, Ghasemi [180]** | | Safety risk management |
| **Elavarasan, Vincent [178]** | | Agriculture |
| **Hossain, Nur [50]**, **Hossain, Nur [182]**, **Hossain, El Amrani [183]** | | Port resilience and management |
| **Chen and Pollino [163]** | | Environmental systems |
| **Phan, Smart [177]** | | Water management |
| **Nikovski [175]** | | Medical diagnosis |
| **Nakatsu [176]** | | Fault diagnosis |
| **Poolsappasit, Dewri [186]**, **Frigault and Wang [187]** | Attack Graphs Modeling | Cyber risk management |
| **Liu and Man [193]**, **Wang and Guo [190]**, **Shin, Son [188]**, **Shin, Son [189]** | Vulnerability assessment | |
| **Wilde [191]**, **Apukhtin [192]** | Insider threat modeling | |

### 3.3 The Integrated Bayesian Network ZT Model Formulation

This article introduces an integrated BN model to examine the feasibility of adopting ZTA within the SMBs' ecosystem. The integrated model comprises two sub-models. The first sub-model is the ZT Implementation Success Model, which quantifies the uncertainties influencing the successful implementation of ZTA pillars and tools while accounting for organizational and financial barriers inherent to SMBs. The second sub-model is the SMBs Risk Analysis Model, which evaluates how ZT adoption impacts the overall magnitude of cyber risk in the presence of common and severe cyberattacks. Figure 3 shows a schematic diagram for the integrated model.

It shows both sub-models' constituents and illustrates how they are integrated. For the ZT Implementation success model on the right side of Figure 3, the relevant ZT security tools that could secure the IT infrastructure for SMBs are selected and connected as a catalyst for the successful implementation of ZT. In addition, financial and organizational barriers are also represented as they negatively impact the success chance of ZTA.

On the left side of Figure 3, the Risk Analysis Model (RAM) is depicted, which considers the most prevalent attack vectors targeting business organizations, the potential targeted assets, and the corresponding ZT controls that could safeguard the assets against the selected attack vectors and reduce the likelihood of data breach incidents resulting from such attacks. The output variable is the risk magnitude, which is dependent on both the successful implementation of ZT components and the potential negative impact of data breaches.

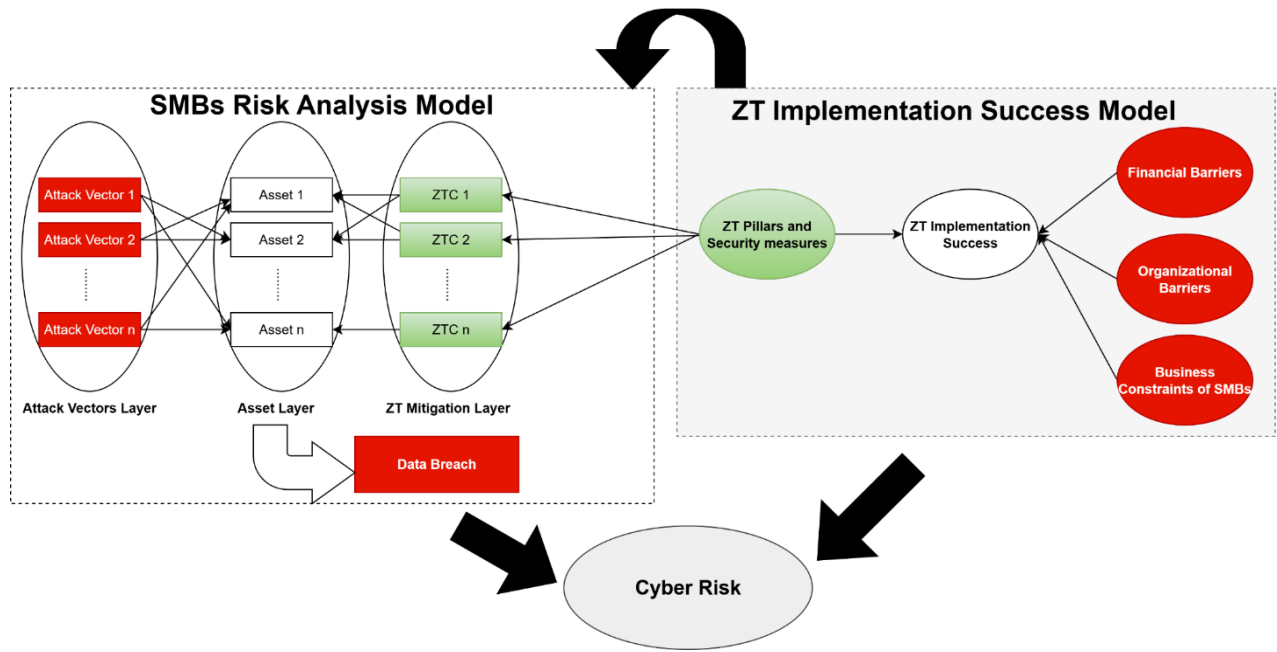

Figure 3 A conceptual framework for the BN ZT integrated model.

This section outlines the integrated BN framework, detailing its underlying assumptions, selected variables, and structural design. It also briefly introduces GeNIE© as the chosen simulation platform, highlighting its analytical capabilities, such as sensitivity and influence analysis, and describing the process of importing and executing the conceptual model within the software environment.

The GeNIE© modeler is an advanced graphical interface developed at the University of Pittsburgh for constructing Bayesian and decision-theoretic models grounded in Bayesian inference theory. Initially created for academic and research purposes, it has since become a widely adopted tool across universities and industries, including government, defense, and commercial sectors, due to its flexibility and analytical reliability [89].

The rationale for selecting GeNIE© in this research lies in its robust analytical capabilities and suitability for academic modeling. GeNIE© allows for advanced sensitivity and influence analysis, effectively identifying the most critical nodes and relationships within a network. It supports both predictive and diagnostic inference, enabling testing of various cybersecurity scenarios and causal relationships under uncertainty. Moreover, as BayesFusion provides a free academic license, the tool offers cost-effective access to powerful modeling and simulation functionalities. These combined advantages make GeNIE© an ideal platform for developing, testing, and analyzing the proposed Bayesian Network models for Zero Trust implementation in SMBs.

3.3.1 Modeling Practices

The proposed BN model adheres to established good modeling practices, as recommended by Jakeman, Letcher [90], and Chen and Pollino [77]. These practices include defining clear objectives, delineating model scope and required resources, creating conceptual models, constructing network structures, determining variable states, and quantifying uncertainty. As suggested by Chockalingam, Pieters [24], node selection is optimized to maintain parsimony and prevent unnecessary network complexity by limiting the number of variables and subdividing large networks into smaller sub-models when needed. Conditional Probability Tables (CPTs) are derived from expert knowledge and historical data, utilizing suitable sampling or learning algorithms, such as Gibbs Sampling, Expectation Maximization, and Gradient Descent, to estimate conditional probabilities even when data are incomplete [77].

Model validation incorporates the three axiom-based method, which measures the coherence of dependencies among nodes, along with sensitivity analysis to evaluate node significance and prioritize cybersecurity measures based on risk. Forward and backward propagation analyses are also leveraged to support decision inference and identify high-impact variables [70].

3.3.2 Model Underlying Assumptions

The model development is guided by specific assumptions to ensure conceptual and contextual alignment with the research objectives:

1. ZT pillars, Identity, Data, Devices, and Applications, and their corresponding relevant controls for SMBs are assumed to be both applicable and achievable within SMBs, given their limited expertise as indicated by Luckett [63].

2. Since no universal definition of SMBs exists globally, this research adopts the U.S. Small Business Administration (SBA) classification, defining SMBs as organizations with up to 1,500 employees.

3. The focus is restricted to U.S.-based cyber incidents documented in the Advisen dataset, due to the higher volume and completeness of event reporting compared to other regions.

4. Given SMBs' early-stage ZT maturity and their primary dependence on organizational and financial readiness, the model concentrates on these two dominant barriers to ZT implementation, while other factors, such as technical or supply chain challenges, are reserved for future investigation as SMBs progress in ZT adoption maturity.

3.3.3 Dataset Preparation and Search Algorithm

The Advisen dataset employed in this research is a large-scale repository encompassing over 137,000 cyber incident records involving 50,206 entities worldwide. It provides a comprehensive view of security incidents such as data breaches, identity theft, and service disruptions, alongside their financial impacts, including recovery costs, economic losses, fines, and penalties. The dataset contains 101 attributes per event, such as case type, incident status, and loss magnitude, with two particularly important fields, case description and proximate cause, that aid in understanding how attacks initiate and unfold, enabling the identification of causal links for BN modeling. Readers are encouraged to refer to the work of Javadnejad, Abdelmagid [22], who proposed a comprehensive exploratory analysis for the Advisen dataset.

To extract incidents relevant to U.S.-based SMBs from the Advisen dataset, a specific text-based search algorithm was developed. The pseudo-code of Algorithm 1 is proposed in Figure A.1, Appendix A. Its objective is to isolate cyber events matching the research scope by linking SMB records from a publicly available list[1] to company names in the Advisen dataset. The algorithm first standardizes company names and country identifiers from both sources to ensure consistent formatting, then filters only U.S. SMBs before cross-referencing these against the Advisen entries. The matching stage identifies incidents reported by those SMBs, providing an accurate dataset for probability estimation and CPT construction. To ensure data accuracy, duplicates were eliminated using the unique incident identifier (MSCAD_ID). All algorithmic steps were coded and executed in the Google Colab environment, selected for its accessible, cloud-based platform and pre-installed Python libraries, eliminating the need for complex local configurations.

Algorithm 1 effectively isolated SMBs' incidents from the broader dataset. From an initial list of 3,365 SMBs, the algorithm identified 2,415 as U.S.-based, while 117,305 of the 137,439 total incidents were tagged with a U.S. country code. After normalization and matching, 1,486 cyber incidents were confirmed as directly involving U.S. SMBs. Verification using the MSCAD_ID field ensured the removal of duplicates, reinforcing the reliability and integrity of the extracted records. The process, executed in just 6.606 seconds on a Dell system with Intel(R) Core(TM) i7-9850H CPU @ 2.60 GHz processor and 16.0 GB RAM, validates the efficiency of the approach. The refined dataset provides the empirical foundation for subsequent stages of Bayesian Network modeling, particularly in estimating prior probabilities and constructing CPTs for different cyberattack types targeting SMBs.

---
[1] https://www.scribd.com/doc/15918817/List-of-SME-Companies

3.3.4 ZT Implementation Success Model Sub-Model

Adopting ZTA introduces a complex array of risks that organizations must be ready to manage, as is the case with any transformative technology. While ZTA can significantly strengthen the cybersecurity posture of SMBs through robust access controls and persistent verification, its implementation presents both organizational and financial challenges, which may impede adoption and introduce new risks, as well as residual risks not fully mitigated by ZT defenses. For SMBs, these risks are compounded by limited budgets, a shortage of technical talent, and operational inefficiencies.

To systematically assess these challenges, this research introduces an integrated model, the Bayesian Network Zero Trust Model (BNZTM), comprised of two sub-models. The first, termed the ZT Implementation Success Model (ZT-ISM), identifies and quantifies the parameters affecting successful ZT deployment in SMBs and evaluates their effect on overall cyber risk. The model is structured to capture both the ZT pillars and their respective security controls, derived from literature and coded in the model, as well as key organizational and financial barriers unique to SMBs. As illustrated in Figure 3, successful ZT adoption is modeled as a function of these pillars and barriers, with the expectation that comprehensive compliance will reduce the magnitude of cyber risk for SMBs.

Selecting and implementing the right ZT pillars is critical for SMB success. Following the guidance of Luckett and others, this research prioritizes the Identity, Devices, Applications, and Data pillars of the CISA ZT maturity model, as these are most suitable for SMB environments in terms of scalability and cost. Larger and more complex pillars (e.g., Network, Visibility, and Automation) can be outsourced to managed security providers until an SMB reaches greater ZT maturity [91]. Table 2 summarizes the considered parameters of ISM and their associated descriptions.

Table 2 The ISM parameters have been considered, along with their associated definitions.

| Parent Node Name | Child Node Name | Description |
|---|---|---|
| Financial Barriers | ZT Costs | Implementing ZTA requires significant investment in upgrading from legacy IT systems to new, advanced technologies. These changes incur overhead and operational expenses [48], often necessitating additional staff to manage the migration and operational complexity [49]. |
| | ZT Budget Estimation | SMBs often face uncertainty in budget planning for ZTA adoption [53]. The absence of standardized definitions and benchmarks, coupled with limited internal understanding, complicates feasibility assessments and makes it difficult |

| Parent Node Name | Child Node Name | Description |
|---|---|---|
| | | to quantify the return on investment for ZTA technologies [50, 52]. |
| | Limited Cybersecurity Budget | SMBs characteristically operate with restricted cybersecurity budgets [92]. Although they are at risk of major cyberattacks just like larger firms, SMBs frequently lack the financial means to invest in robust defense measures. The expense of adopting and maintaining advanced ZTA solutions poses a significant barrier [38]. |
| Organizational Barriers | Legacy Systems Substitution | Transitioning to ZTA often involves replacing or integrating with existing security systems, particularly those based on OT. For SMBs, less complex infrastructures can make system substitution easier, but compatibility issues may still arise. |
| | Resistance to Change | Employee reluctance can impede the move to ZT frameworks, as new policies and stricter access controls are sometimes viewed as unnecessary disruptions. Organizational culture and employee acceptance are essential for a successful migration [52]. |
| | Analysis Paralysis | Decision-makers in SMBs can face analysis paralysis due to the unfamiliarity and perceived complexity of ZTA. The multiple new technologies and policies involved may overwhelm leaders, especially if they lack strong cybersecurity knowledge or risk-management experience. |
| | Hiring A Security Analyst | Effective ZTA deployment demands skilled personnel, especially security analysts, who can administer policies, conduct risk assessments, and oversee migration phases [64]. However, SMBs may struggle to hire such experts due to financial limitations, making the availability of security talent largely dependent on available budget [93]. |
| ZT Relevant Pillars for SMBs and their associated security measures | Identity Pillar | The Identity pillar centers on precise control of user or entity access to organizational resources through robust authentication, such as Multi-Factor Authentication (MFA), Single Sign-On (SSO), and Role-Based Access Control (RBAC) [91]. These measures help enforce the principle of least privilege, enhancing trust and network visibility [63]. |
| | Data Pillar | This pillar involves protecting organizational data in all forms, at rest, in transit, and in use, through mechanisms like data encryption and Data Loss Prevention (DLP) systems [63]. Affordable, cloud-based solutions can be used by SMBs to secure sensitive information, promote compliance, and mitigate breach risks [94]. |

| Parent Node Name | Child Node Name | Description |
|---|---|---|
| | Devices Pillar | The Devices pillar emphasizes inventory and control of all network-connected assets, such as laptops, IoT devices, and mobile endpoints [91]. Creating and maintaining asset inventories, continuous monitoring via Endpoint Detection and Response (EDR), and employing Device Management tools like Unified Endpoint Management (UEM) and Mobile Device Management (MDM) are key strategies, with cost-effective options available for SMBs [63]. |
| | Applications Pillar | Application security safeguards systems, apps, and digital services by enforcing policy-based or role-based access controls and utilizing tools like Web Application Firewalls (WAFs). While Policy-Based Access Control (PBAC) offers fine-grained management, Role-Based Access Control is often favored in SMBs for its simplicity and practicality [95]. Protecting data transfers between applications via secure APIs is also essential [63, 65, 96]. |

3.3.5 ZT Risk Analysis Model Sub-Model

The SMBs Risk Analysis Model (RAM) is designed to evaluate the effectiveness of selected ZT security controls in reducing overall cyber risk for small and medium-sized businesses (SMBs). The model begins by identifying the most significant and frequently occurring attack vectors that impact SMBs, with consideration for both frequency and severity to accurately define risk magnitude as a product of likelihood and consequence [97].

The next step involves determining the organization's critical assets, those most susceptible to targeting by malicious actors. Recognizing these vital assets enables SMBs to adopt a risk-based ZT strategy, ensuring priority protection for essential resources and optimizing security investments. Asset criticality assessment is a cornerstone of effective ZT implementation and risk management [98]. Research and recent industry reports, such as the 2024 Verizon Data Breach Investigations Report (DBIR) and the 2024 IBM Cost of a Data Breach Report [11, 99], are referenced extensively in the literature to inform the mapping of prevalent cyber threats and their impact on SMBs. These reports serve as well-established benchmarks for understanding cyberattack trends, consequences, and risk reduction benefits associated with robust ZT practices, including notable reductions in breach costs.

Causal relationships between attack vectors and target assets are modeled using data insights from industry reports, the Advisen cyber incident dataset, and advanced Generative AI tools for enhanced causal reasoning and alignment with organizational objectives [100]. Literature supports the application of large language models (LLMs) for interpreting and correlating cyber threats to attack patterns and tactics, as indicated by Fayyazi and Yang [101].

3.3.5.1 Targeted Assets

SMBs commonly rely on a range of technology-enabled systems and devices, such as computers, for fundamental business operations. These systems underpin activities including customer data management, financial transactions, supply chain oversight, and various other mission-critical processes. The occurrence of cyberattacks threatens to significantly disrupt these essential functions, potentially resulting in financial losses, breaches of sensitive information, and erosion of customer trust. A detailed overview of each key asset and its role within SMBs is outlined in the following bullet points.

- **Servers** play a crucial role in maintaining business continuity for SMBs, facilitating essential functions without direct end-user interaction [99]. They can be either hardware or software, supporting various phases of operations by providing data storage, transfer capabilities, and advanced software solutions that enhance both customer experience and organizational productivity. Four primary types serve distinct functions: web servers deliver website content and manage internet traffic; file servers store and enable access to shared documents; database servers manage data applications and facilitate secure, efficient data retrieval; and application servers support the operation of business-specific programs [102]. SMBs commonly employ servers for hosting email domains, storing business files, providing VPN-based remote access, handling user authentication, hosting websites, running applications for HR and accounting, and performing data backups [103]. Given their integral role and the vast data they process, servers are prominent targets for cyberattacks, evidenced by the 2024 DBIR, which attributes over 80% of 8,190 reported breaches to server compromise [99]. Therefore, it is vital to safeguard servers with firewalls, multi-factor authentication (MFA), and Zero Trust encryption to ensure the integrity and confidentiality of organizational data.
- **Emails and Websites** are vital digital assets for SMBs, serving as core channels for daily operations and stakeholder engagement. Websites not only present products and services but also function as communication gateways with customers. Likewise, email facilitates internal and external business correspondence. These assets often store or transmit sensitive information, making them vulnerable to defacement, phishing, or insider threats. Attacks on websites can result in misinformation and reputational harm, while email breaches may disclose confidential data. Strong access controls and MFA, as prescribed by ZT tools, can help mitigate risks posed by techniques like social engineering and insider misuse.
- **Printed Records** are physical data assets containing sensitive information such as customer PII, making them attractive to adversaries. Unauthorized disclosure, whether accidental or resulting from insider threats, can cause reputational and financial damage. Enforcing least privilege determines access based on necessity, thereby limiting unauthorized data exposure [104]. Physical security controls, such as role-based access for printing and document handling, alongside clear verification of personnel identities, are key to securing printed records [105].
- **User Devices**, including laptops, desktops, smartphones, and tablets, are central to SMB operations, supporting appointments, supply chain communications, and remote system monitoring [99]. These assets typically hold critical data, such as customer details and financial records. Loss or compromise of user devices can lead to severe legal, financial, and reputational consequences. Data breaches often result from device theft or accidental exposure. Recommended ZT measures include Endpoint Detection and Response (EDR) for anomaly detection, Data Encryption (DE), and RBAC for physical and digital access control. Regular device monitoring and compliance with CISA guidelines are essential to prevent vulnerabilities and exploitation [106].
- **Portable Data Storage Tools** such as external hard drives, flash drives, and memory cards provide a convenient means for storing and transferring sensitive information [107]. However, because these devices are easily lost or stolen, they require the deployment of data encryption and Data Loss Prevention (DLP) tools to ensure data confidentiality and sustainable backups [108].
- **Software** includes all applications, system programs, or drivers necessary for device and business operations [109]. Safeguarding software from unauthorized access or destruction mitigates common threats such as malware and ransomware, which are highly prevalent in SMB environments [22, 110]. Security measures within the Application ZT pillar, like web application firewalls and robust identity management (RBAC), are essential to oversee, filter, and block malicious activity, protecting applications and sensitive data from compromise [94].

3.3.5.2 The Modelled Attack Vectors

A cyberattack vector is defined as the pathway or method used by attackers to infiltrate a network or system, typically by exploiting vulnerabilities within the organization's cybersecurity infrastructure [111]. It encompasses the techniques malicious agents employ to exfiltrate sensitive information across different digital assets [112]. The identification and assessment of attack vectors in this research commenced with an analysis of industry reports, evaluating each vector based on risk components such as frequency and severity, as depicted in Figure 4. To further confirm their relevance, the chosen attack vectors were cross-referenced with real-world incidents documented in the Advisen dataset to demonstrate their prevalence among SMBs. Supporting evidence from contemporary literature also reinforced the selection of these attack vectors.

This approach ensures greater confidence in the relevance of threats to SMBs, accounting for the fact that SMBs often face the same cyberattack types as large enterprises, a pattern widely recognized in data breach and cybersecurity reports [113]. Chidukwani, Zander [5] conducted a literature review focusing specifically on cyber risks and the implementation challenges experienced by SMBs, revealing broad agreement that the most common attacks include social engineering (especially phishing), insider threats, and credential-based attacks such as identity theft. Fernandez De Arroyabe and Fernandez de Arroyabe [114] highlighted phishing, asset misuse, insider threats, and unauthorized use of devices, whether accidental or intentional, as key contributors to data breaches in SMBs. These attack vectors are consistently identified in leading industry reports as major threats causing significant financial impacts. In this context, a data breach refers strictly to incidents where attackers gain unauthorized access to confidential information [115].

- **Phishing** attacks employ diverse techniques, including targeted spear phishing and broad malware spam campaigns. They often use malicious links, third-party platforms like social media, or sender identity spoofing to deceive recipients and bypass security measures [116]. These attacks typically serve as the initial step towards achieving malicious goals, with phishing being the most frequent social engineering method associated with identity theft [63, 113]. According to CISA, phishing is responsible for over 90% of successful cyberattacks [117]. SMBs are particularly susceptible due to limited cybersecurity awareness and weaker defenses compared to larger organizations. Given SMBs' critical roles in sectors like energy and healthcare, phishing attacks can disrupt operations and compromise sensitive information, sometimes triggering cascading attacks that affect larger entities [65]. Industry reports such as IBM and Verizon underscore phishing as a leading cause of data breaches, accounting for 31% of social engineering attacks primarily via email, with significant financial losses averaging $4.88 million [11].
- **Insider threats** comprise malicious, negligent, and compromised insiders, along with third-party risks. Malicious insiders, including disgruntled employees or former staff, exploit access for revenge or financial gain, sometimes collaborating with external adversaries [118]. Negligent insiders cause security incidents unintentionally, for example, by misplacing devices containing sensitive data. The 2024 Ponemon report cites negligent users as the leading cause of data loss incidents at 70.6% [119]. Compromised insiders have stolen credentials used to breach systems [118]. Insider threats affect organizations across all sizes and sectors, with insiders typically holding privileged access that can cause severe damage, especially in critical industries like healthcare [99]. Although malicious insider incidents are less frequent (7%), they cause the highest financial impact, averaging $4.99 million per incident. IAM tools, vital for combating insider threats, manage identities and access rights to ensure authorized use of resources [118].

- **Physically Lost or Stolen Assets**, such as user devices and printed records, poses significant risks by enabling unauthorized access to sensitive data [120]. Physical breaches often result in data theft, given the ease of extracting information from stolen devices compared to penetrating network defenses [121]. Despite their lower frequency, physical security incidents led to financial losses averaging $4.19 million in 2024, illustrating that infrequent attacks can be highly damaging [114].

- **Credential-Based Attacks** begin with the theft of access credentials, commonly via phishing or brute-force methods [99, 122]. Phishing is a prevalent and cost-effective attack vector that exploits human error and inadequate awareness [122]. The 2024 DBIR reports that 14% of credential-based attacks start with phishing and that stolen credentials were leveraged in 77% of web application breaches [99]. From 2013 onwards, credential-based attacks accounted for 31% of data breaches among nearly 36,000 incidents [99]. IBM ranks these attacks as the most frequent, with a 16% occurrence rate and average financial losses of $4.81 million. Attackers prioritize critical data such as customer personally identifiable information (PII), intellectual property, and employee data [11]. Key targeted assets include web applications and email systems, where stolen credentials enable unauthorized access, data theft, and privilege escalation. Mitigating these risks involves deploying MFA and Web Application Firewalls (WAF) as part of Zero Trust security tools.

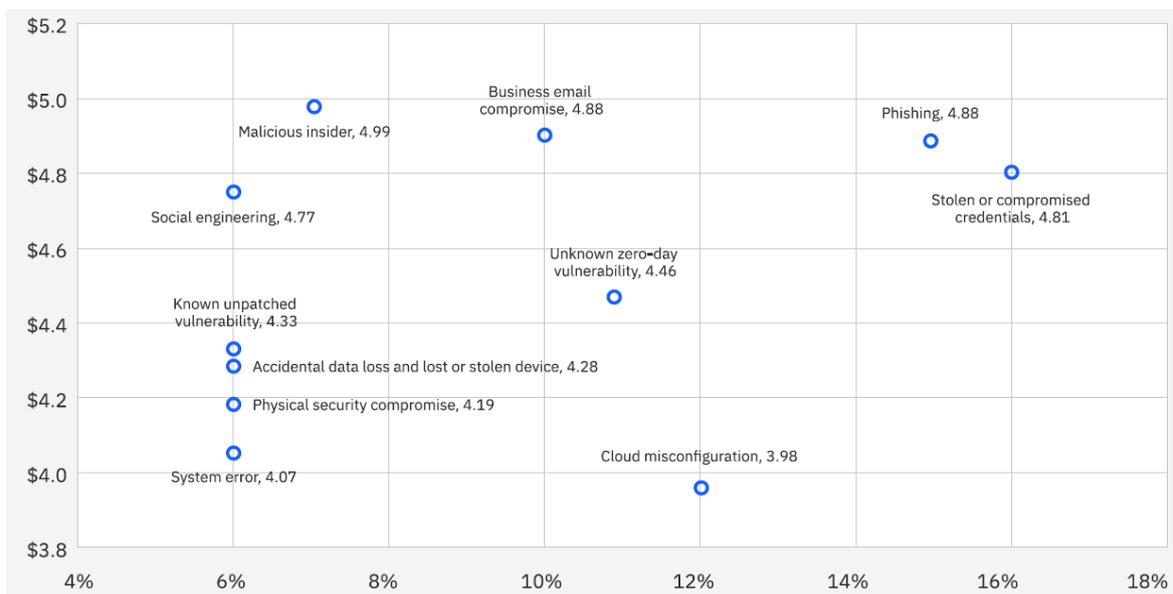

Figure 4 Cost and frequency of data breach incidents based on various initial attack vectors [65].

### 3.3.5.3 ZT Security Controls

- ZTC1_Data

The purpose of ZTC1, which refers to Data controls, is to safeguard sensitive information processed through the digital assets of SMBs by utilizing security tools such as Data Loss Prevention (DLP) and Data Encryption (DE). Data encryption protects against unauthorized access by converting data into an unreadable format for anyone without the decryption key, ensuring security for data in transit, at rest, or being processed. Data backup and recovery involve creating secure copies of data to preserve business continuity in the event of loss or corruption due to unforeseen circumstances [123]. These encryption tools are especially critical for protecting information stored on end-user devices and portable storage media [99]. DLP protects against both accidental and intentional data disclosures, representing a crucial success factor within the Data pillar by preventing data leakage and maintaining adequate levels of data security [94]. Studies by Luckett [63] and emphasize that DLP and DE are affordable and vital for SMBs adopting ZTA, with bundled services available from providers such as Palo Alto and Cloudflare. These measures align with SMBs' growing reliance on data analytics to remain competitive, making robust data protection indispensable for preventing breaches.

- ZTC2_Devices

  Protecting organizational devices starts with identifying all equipment connected to the network, including smartphones, laptops, BYOD, and IoT devices, through comprehensive device inventories. This inventory establishes visibility and defines security baselines necessary for effective monitoring and access control [124]. Luckett [63] underscores the importance of asset identification for successful ZT implementation regardless of organizational size. Device management involves using technologies such as Unified Endpoint Management (UEM) and Mobile Device Management (MDM) to secure a wide range of IT devices, including computers and mobile phones [124]. These tools can be deployed cost-effectively, through providers such as JumpCloud which offers IAM bundled with device management for less than $20 per user per month [63]. Endpoint Detection and Response (EDR) systems offer essential protection by detecting malicious or anomalous activities, preventing potentially harmful outcomes like malware infections triggered by phishing [125].

- ZTC3_Identity

  ZT identity controls focus on authenticating and authorizing access for users, devices, and applications, forming the foundation for access control decisions handled by the ZT policy engine [123]. Multi-Factor Authentication (MFA) is a key security measure, leveraging multiple verification sources to drastically reduce unauthorized access risks [94]. MFA is cited by Luckett [63] as essential across organizations of all sizes due to its affordability and ease of use, significantly mitigating the risk of credential exfiltration. Single Sign-On (SSO), recommended alongside MFA, simplifies credential management by reducing login prompts and easing MFA integration across applications, making it accessible for SMBs [63]. Role-Based Access Control (RBAC) restricts access according to user roles, limiting data exposure by granting only necessary permissions, for instance, allowing a "Research Scientist" access to R&D materials but not a "procurement officer" [126]. RBAC protects both digital assets and physical resources, reducing breach risks and data leakage.

- ZTC4_Application

  For SMBs, a cost-effective and suitable ZT security solution for the application layer is Web Application Firewall (WAF) technology [63, 113]. WAFs protect Application Programming Interfaces (APIs), which facilitate data exchanges among applications containing sensitive information [113]. By monitoring and filtering malicious traffic at the application layer, WAFs help defend against cyberattacks such as credential theft,

Cross-Site Scripting (XSS), SQL injection, and data exfiltration attempts [127, 128]. Table 3 consolidates the cyber threats, targeted assets, and corresponding ZT security controls developed to mitigate these risks effectively.

The integrated Bayesian Network Zero Trust Model (BNZTM) links specific ZTCs to targeted assets to defend against identified cyberattacks, with the risk magnitude serving as the response variable. Within the Risk Analysis Model (RAM), the selected attack vectors are recognized for their potential to cause data breaches by exposing sensitive information. Such breaches can inflict substantial harm on SMBs, particularly in sectors like healthcare, finance, and the public sector, where regulatory penalties can exacerbate financial losses [115]. Data breaches are regarded as catalysts for increasing organizational cyber risk due to direct and indirect repercussions, including legal, reputational, and operational damages. The mapping of ZT tools, aligned with the Implementation Success Model (ISM) and depicted in Figure 3 through ZT control nodes, follows CISA's Zero Trust Maturity Model (ZTMM) recommendations. These controls primarily aim to restrict unauthorized data access, the principal cause of breaches. Consequently, the efficacy of ZTCs significantly influences breach frequency and severity, thereby impacting the overall cyber risk faced by SMBs, underpinning the causal link between data breach events and cyber risk in the model as illustrated in Figure 3.

Table 3 The identified prevalent attack vectors, critical assets, and the corresponding ZT security control(s).

| Attack Vector | Targeted Asset | | | | | | |
|---|---|---|---|---|---|---|---|
| | Server | Email | Website | Printed Records | User Devices | Software | Portable Data Storage Instruments |
| Phishing | √ | √ | | | | | |
| Insider Threat | √ | √ | √ | √ | √ | √ | √ |
| Physically Lost or Stolen | √ | | | √ | √ | | √ |
| Credential-Based Attacks | √ | √ | √ | √ | √ | | √ |
| Potential ZTC(s) | ZTC1, ZTC3 | ZTC2, ZTC3 | ZTC1, ZTC3 | ZTC3 | ZTC1, ZTC2, ZTC3 | ZTC3, ZTC4 | ZTC1 |

## 3.4 Quantifying BN-ISM Parameters

In Bayesian statistics, a prior distribution represents the initial knowledge or belief about an unknown parameter before incorporating new data. Priors are generally categorized into three types based on the certainty about the parameter: non-informative (or weakly informative), informative, and diffuse [129, 130]. Non-informative priors have minimal impact on the posterior distribution, serving as neutral inputs, whereas informative priors incorporate substantial prior knowledge derived from empirical evidence like expert opinion or industry data [130], providing credible probability distributions, especially useful when data are sparse. These informative priors help stabilize parameter estimates by effectively integrating both prior knowledge and observed data.

Parameters in a BN correspond to nodes representing variables, which may be categorical, Boolean (discrete), or continuous [77]. Boolean nodes have two mutually exclusive states, usually true or false, with probabilities that complement each other. Continuous nodes represent variables with potentially infinite values modeled by appropriate probability distributions. Choosing suitable informative prior distributions depends on the parameter type and mathematical properties such as conjugacy, where the posterior distribution mirrors the same family as the prior, simplifying computation [130]. For binary outcomes modeled by Bernoulli or Binomial likelihoods, the Beta distribution serves as a conjugate prior [130]. This study applies informative priors to the Boolean parent nodes in the ZT Implementation Success Model (BN-ISM), with the rationale and mathematical formulation for prior probability estimation detailed in subsequent sections.

### 3.4.1 Informative Prior Probability Estimation for ISM parameters

Mathematically, an informative prior for a parameter $\theta$ is described as a probability distribution $P(\theta)$ over the parameter space, where $\theta \in \Theta$. In the context of a Binomial distribution model, data consist of a fixed number of trials, each resulting in one of two possible outcomes: "success" or "failure." The data are summarized by the total number of successes observed across these $n$ trials, denoted by $y$. The binomial sampling model for this setup is presented in Equation 6.

$$P(y|\theta) = \binom{n}{y} \theta^y (1-\theta)^{n-y} \qquad \text{(Equation 6)}$$

The likelihood function is then expressed as,

$$P(y|\theta) \propto \theta^a (1-\theta)^b \qquad \text{(Equation 7)}$$

An informative prior distribution can be mathematically expressed using a specific form, such as the Beta distribution, characterized by parameters $a$ and $b$, which act as hyperparameters governing the distribution's shape. When the prior and posterior distributions share the same functional form but differ in these parameters, the conjugacy simplifies Bayesian updating. The prior density function, represented by the Beta distribution with parameters $\alpha$ and $\beta$, corresponds to prior successes and failures, as outlined in Equation 8. The posterior density function for the parameter is similarly derived and shown in Equation 9.

$$P(\theta) \propto \theta^{\alpha-1}(1-\theta)^{\beta-1} \qquad \text{(Equation 8)}$$

$$\begin{aligned}P(\theta) &\propto \theta^y(1-\theta)^{n-y}\theta^{\alpha-1}(1-\theta)^{\beta-1}\\ &= Beta(\theta|\alpha+y, \beta+n-y)\end{aligned} \qquad \text{(Equation 9)}$$

The informative prior's key properties are captured by its moments, mean, variance, and Effective Sample Size (ESS), which elucidate the central belief about the parameter and the strength of prior evidence [130]. The prior mean indicates the expected parameter value before observing new data and is computed based on the Beta distribution's hyperparameters $\alpha$ and $\beta$ (Equation 10).

$$E[\theta] = \frac{\alpha}{\alpha + \beta} \qquad \text{(Equation 10)}$$

The variance quantifies the uncertainty around this mean value, with a smaller variance reflecting stronger confidence (Equation 11). ESS estimates the effective amount of information the prior encodes, analogous to a sample size, and is derived as shown in Equation 12.

$$Var[\theta] = \frac{\alpha\beta}{(\alpha + \beta)^2(\alpha + \beta + 1)} \qquad \text{(Equation 11)}$$

$$ESS = \alpha + \beta \qquad \text{(Equation 12)}$$

This study employs informative prior estimation grounded in numerical data gathered from industry sources such as the 2020 Untangle SMB IT security report [92]. Given that ISM variables are Boolean (true/false), the Beta distribution is the appropriate conjugate prior. For example, according to Untangle's survey, 32% of 500 SMB respondents identified "*LimitedBudgets*" as a key organizational barrier; this external evidence is used to parameterize the Beta prior via Laplace smoothing, a Bayesian technique for proportion estimation under uncertainty, detailed in Equations 13 to 15. Thus, the informative prior is $Beta\ (161, 341)$.

$$x = pn = 0.32 * 500 = 160 \qquad \text{(Equation 13)}$$
$$\alpha = x + 1 = 160 + 1 = 161 \qquad \text{(Equation 14)}$$
$$\beta = (n - x) + 1 = (500 - 160) + 1 = 341 \qquad \text{(Equation 15)}$$

Subsequently, the prior mean, variance, and ESS are calculated from these parameters as shown in Equations 16, 17, and 18, respectively. The prior mean aligns closely with survey data (32%), the variance indicates strong confidence in this estimate, and the ESS quantifies the equivalent sample size of prior information, highlighting its influence relative to new data.

$$E[\theta] = \frac{161}{161 + 341} = 0.3207 \qquad \text{(Equation 16)}$$

$$Var[\theta] = \frac{160 * 341}{(160 + 341)^2(160 + 341 + 1)} = 0.00043 \qquad \text{(Equation 17)}$$

$$ESS = 161 + 341 = 502 \qquad \text{(Equation 18)}$$

Similarly, the findings of the Untangle report have been employed to determine the prior probability of the variables shown in Table 4. The results of the survey reveal that 24 percent of employees are not willing to follow cybersecurity policies and standards, which could be mapped to the "*ResistancetoChange*" variable. Additionally, 14 percent and 12 percent are the percentages of SMBs that indicate limited cybersecurity awareness and shortages in human resources, respectively. The limited knowledge of robust cybersecurity solutions, such as ZTA, by SMBs' management is found to be 11 percent. Another industry survey conducted by Techaisle 2025 research [131] highlighted that 67% of surveyed SMBs indicated that the top barrier to implementing ZT solutions is the potential high costs. Thus, the published band proportion $p$ is treated as the prior mean and converted to a Beta prior using a small sample size $k = 30$. The value of $k$ is selected to maintain moderate uncertainty, allowing new data to update the posterior.

Additionally, the aggregate percentage of respondents highlighted that the technology gaps between existing IT systems and ZT security requirements and measures are 49 percent. Therefore, the prior probability of the state "Yes" for the Legacy System Substitution variable is 0.49 if the respondents reported this factor as moderate, sizable, or major, and "No" otherwise. Following the same approach discussed in Equations 14 − 18, and with the same moderate sample size $k = 100$, $\alpha = 50$, $\beta = 52$, $K = 102$, $Mean = 0.49$, $Variance = 0.000497$.

Likewise, the Ponemon report [132] indicated that 77 percent of SMBs cited insufficient personnel as a key security limitation, in addition to the fact that only 37 percent of SMBs depend on internal staff to manage their IT operations, with MSSPs covering an additional 32 percent. Hence, this implies that about one-third of SMBs have a dedicated analyst for cybersecurity, leaving about two-thirds lacking dedicated staff that can oversee ZTA implementation. A reasonable informative prior for the probability of *NoHiring* in SMBs is 0.68 with a moderate informative prior of Beta(34, 16). The mean and SD can be calculated as previously, and the values are 0.68 and SD 0.07.

Finally, the aggregate percentage of respondents who indicated that cost concerns are a factor when it comes to implementing ZT security tools is 46 percent. Thus, ZT budget estimation can be inferred with a probability of success equal to 46 percent. Equations 14 − 18 are used to determine the Beta prior distribution for both variables using the same moderate sample size (ESS) $k = 100$. The Beta prior distribution hyperparameters are specified in Table 4. It is worth noting that the Beta distribution hyperparameters' values will not be used in the GeNIE© software while modeling the BN model, since there is no data available for parameters to learn from. The values of $\alpha$, $\beta$, $and\ K$ are reported to enable reproducibility and transparency about the followed estimation approach.

Table 4 Prior Probabilities of financial and organizational barriers to ZT implementation variables belong to the ISM.

| Variable Name | Modeling Approach | Informative Prior Probability Value | | | |
|---|---|---|---|---|---|
| | | p | α | β | Prior Distribution |
| Limited Cybersecurity Budgets | Boolean | 0.32 | 161 | 341 | Beta (161, 341) |
| Resistance to Change | Boolean | 0.24 | 121 | 381 | Beta (121, 381) |
| Cybersecurity Awareness Level | Boolean | 0.14 | 71 | 431 | Beta (71, 431) |
| No Hiring a Security Analyst | Boolean | 0.68 | 34 | 16 | Beta (34, 16) |
| ZT Tech Knowledge Level | Boolean | 0.11 | 56 | 446 | Beta (56, 446) |
| ZT Costs | Boolean | 0.67 | 21.1 | 10.9 | Beta (21.1, 10.9) |

| Variable Name | Modeling Approach | Informative Prior Probability Value | | | Prior Distribution |
|---|---|---|---|---|---|
| | | p | α | β | |
| Legacy Systems Substitution | Boolean | 0.49 | 50 | 52 | Beta (50, 52) |
| Poor ZT Budget Estimation | Boolean | 0.46 | 47 | 52 | Beta (47, 52) |

To estimate the prior probability distributions for ZT security measures, industry reports and surveys were used to obtain relevant information about the adoption rates of ZT measures in the context of SMBs.

Starting with the Identity Pillar, A recent cyber readiness report published by the Cyber Readiness Institute (CRI) specified that only 35% of 2300 SMBs around the world adopted MFA within their digital infrastructure [133]. Thus, this industry report is considered to estimate the prior probability of the MFA variable with a probability value $p = 0.35$. To reflect uncertainty about the representativeness and avoid overstating certainty in downstream nodes, the variable is encoded as a Beta prior with $k = 30$. By substituting these values in Equations 14 − 15, it yields the prior probability distribution of Beta($\alpha = 11.5, \beta = 20.5$).

For the RBAC security measure, reliable SMB-specific adoption data is absent. Three sources from the market were used to estimate the prior probability of RBAC adoption in SMBs. First, Allied Market Research reported that the increasing annual growth rate of SMBs shows an expected rise in the adoption rate [134]. Thus, it can be assumed that a conservative prior mean of $p = 0.51$, and $k = 20$. From Equations 14 − 15, the Beta hyperparameters' values are $\alpha = 10.2, \beta = 9.8$.

For Single Sign-On (SSO), there is no robust quantitative evidence specifically describing its adoption probability among SMBs. A recent CISA report [135] characterizes SSO adoption in SMBs as "low" due to various barriers, while other sources, such as LastPass [136] and Okta [43] provide numerical adoption rates not related to SMBs. Consequently, a weakly informative Beta prior with a small effective sample size is used, with hyperparameters $\alpha = 2$, and $\beta = 8$, reflecting high epistemic uncertainty about SSO uptake and following the weakly informative prior approach of Gelman, Jakulin [137], which avoids strong assumptions while ruling out implausible values.

For the Device pillar, Okta reports that 18% of respondents in the 500 – 999 employee range and 16% in the 1,000 - 4,999 range use device management, and this study assumes roughly equal representation for SMBs under the SBA's ≤1,500-employee definition. A simple average is used as the base estimate (Equation 19), and the corresponding Beta prior is parameterized with the same effective sample size (ESS = 50), yielding $\alpha = 8.5, \beta = 41.5$, without Laplace smoothing.

$$p = \frac{0.18 + 0.16}{2} = 0.17 \qquad \text{(Equation 19)}$$

For Endpoint Detection and Response (EDR), UK and vendor surveys show that 33% of all businesses and 63% of medium-sized firms use security monitoring tools, with SMBs often relying on traditional methods rather than full EDR [138, 139]. A conservative prior mean of 0.40 is adopted, leading to a Beta prior with $\alpha = 20, \beta = 30$, ESS = 50. For Device Inventory, multiple sources, as specified by the UK cybersecurity breaches survey [139] and a vendor survey [140], indicate visibility gaps among SMBs but a substantial minority, especially medium-sized firms, maintain some form of inventory. A mid-range conservative point estimate of 0.50 is used, resulting in a Beta prior with $\alpha = 22.5, \beta = 27.5$, ESS = 50.

Within the Data pillar, Data Encryption (DE) and Data Loss Prevention (DLP) are modeled as separate nodes. Industry surveys suggest that only 17–30% of SMBs have systematic encryption deployed, with Thales reporting limited comprehensive cloud encryption among SMBs [141, 142]. A conservative prior mean of 0.30 is selected for DE involving those with partial encryption and coverage, and applying Equations 14–15 with ESS = 50 yields Beta hyperparameters with $\alpha = 15, \beta = 35$. DLP adoption in SMBs remains low due to financial and technical barriers, though growth is stronger in large and regulated organizations [143, 144]. A conservative prior mean of 0.15 is used, giving a Beta prior with $\alpha = 7.5, \beta = 42.5$.

For the Application pillar, the WAF node is assigned a prior mean of 0.20, consistent with evidence of rapid WAF market growth but lower penetration among SMBs, despite recommendations in the literature [113]. Using the same ESS ($k = 50$), and Equations 14–15, the WAF prior is parameterized as Beta(10, 40). Table 5 consolidates these priors and their associated Beta hyperparameters.

Table 5 Prior Probabilities of ZT security measures variables.

| ZT Pillars Child Nodes | ZT Security Measures Parent Nodes | Modeling Approach | Prior Probability Value and Distribution | | | |
|---|---|---|---|---|---|---|
| | | | $p$ | $\alpha$ | $\beta$ | Prior Distribution |
| Data | DE | Boolean | 0.30 | 15 | 35 | Beta (15, 35) |
| | DLP | Boolean | 0.15 | 7.5 | 42.5 | Beta (7.5, 42.5) |
| Identity | SSO | Boolean | 0.20 | 2 | 8 | Beta (2, 8) |
| | RBAC | Boolean | 0.51 | 10.2 | 9.8 | Beta (10.2, 9.8) |
| | MFA | Boolean | 0.35 | 11.5 | 20.5 | Beta (11.5, 20.5) |
| Device | DM | Boolean | 0.17 | 8.5 | 41.5 | Beta (8.5, 41.5) |
| | D_INV | Boolean | 0.45 | 22.5 | 27.5 | Beta (22.5, 27.5) |
| | EDR | Boolean | 0.40 | 20 | 30 | Beta (20, 30) |
| Application | WAF | Boolean | 0.20 | 10 | 40 | Beta (10, 40) |

3.4.2 Causal Links Strength Values for the BN-ISM Nodes

In the BN-ISM, the conditional probability distributions of child nodes are estimated using the *NoisyOR* function, which efficiently models how multiple parent nodes independently influence a binary child variable. Direct elicitation of full conditional probability tables becomes impractical and error-prone when many parents are involved, because it would require specifying probabilities for all combinations of parent states; the *NoisyOR* formulation greatly simplifies this task. Originally introduced as a precursor to the *Noisy-Max* model for binary variables, *NoisyOR* treats each parent as an independent causal factor with an associated strength parameter and includes a "leak" term to capture unmodeled causes, i.e., the probability that the child is true even when all modeled parents are false.

Suppose $X_1, \ldots, X_n$ be $n$ Boolean variables. For each $i = 1, \ldots, n$ let $v_i$ be a number $\in [0, 1]$, where $v_i$ is the assigned weight (e.g., strength) to $X_i$. This value captures the probability that the consequence will be true if that certain parameter (and no other parameter) is true. Let $l$ be a number $\in [0, 1]$, where it represents the leak factor, which represents the variables not considered in the model. In other words, the leak value is the probability of the child node variable being true if all the linked parent nodes are absent [145].

Let $Y$ be a Boolean variable with parents $X_1, \ldots, X_n$, then

$$Y = NoisyOR(X_1, v_1, X_2, v_2, \ldots \ldots \ldots \ldots, X_n, v_n, l) \qquad \text{(Equation 20)}$$

and the conditional probability can be estimated from the following expression (Equation 21):

$$P(Y = true \,|X_1, \ldots \ldots, X_n) = 1 - (1 - l) \times \prod_{X_i \text{ is true}}(1 - v_i) \quad \text{(Equation 21)}$$

In practice, each parent is assigned a causal weight and a small leak probability, and the child's conditional probability is then computed for each configuration of parent states using the *NoisyOR* formula shown in Equations 20 - 21, thereby reducing the complexity of probability elicitation while preserving a plausible causal interpretation.

To estimate the strength values for the causes of the child nodes, each parent node's influence on a child node is quantified by first assigning that parent a posterior Beta distribution, derived from industry reports and surveys, and summarized in Table 4. These Beta distributions encode uncertainty about the parents' marginal probabilities, but because *NoisyOR* causal strengths cannot be inferred directly from marginals, a prior predictive procedure is applied to obtain plausible ranges for the causal weights. The procedure proceeds as follows:

1. For every parent, Monte Carlo samples are drawn from its posterior Beta distribution to approximate how frequently the child node would be active in the population.

2. For each parent–child edge, a Beta prior is then specified for the causal strength, using either weakly informative assumptions or literature-based values with a conservative effective sample size so that the prior mean is neutral and the variance remains large; 20,000 Monte Carlo draws are generated for these strengths.

3. The leak parameter is held fixed (for example, Beta(2,48)), unless otherwise justified, to represent the probability that the child activates when all modeled parents are absent.

4. In each Monte Carlo iteration, one sample is drawn for the parent marginal, one for the causal strength, and one for the leak; the implied child probability under each parent configuration is then computed using the NoisyOR equation.

5. The resulting empirical distribution of strength values captures uncertainty for each causal link, from which the mean, median, and 95% credible interval are obtained; the posterior median for each link is finally used as the point estimate entered into GeNIE.

For the FinancialBarriers node, the causal influence values of its three parents, *LimitedBudget*, *ZTCosts*, and *ZTBudgetEstimation* are encoded using Beta(14, 6), Beta(6.5, 3.5), and Beta(2, 8), respectively. The first prior implies a relatively high expected causal strength (mean 0.7) for limited budgets, while still allowing substantial uncertainty through a wide credible interval. *ZTCosts* is modeled with a mean of 0.65 and a variance of 0.02, reflecting that ZT-related expenses are generally impactful but vary with factors such as vendor choice and deployment scope. *ZTBudgetEstimation* is treated as less influential, with a mean of 0.2. As reported in Table 6, the assigned means closely match the expected counterparts for the parent parameters. The leak term is set to Beta(1, 24), with a small mean of 0.04, indicating that most financial barriers are accounted for by the modeled causes and that residual, unmodeled causes are rare.

Table 6 Result of MC simulations of the FinancialBarriers node.

| Variable Name | Mean | Median | SD | $q_{2.5\%}$ | $q_{97.5\%}$ |
|---|---|---|---|---|---|
| LimitedBudget | 0.6999 | 0.7069 | 0.1006 | 0.4868 | 0.8743 |
| ZTCosts | 0.6483 | 0.6577 | 0.1447 | 0.3480 | 0.8974 |
| PoorZTBudgetEst | 0.1995 | 0.1791 | 0.1194 | 0.0291 | 0.4788 |
| Leak | 0.0400 | 0.0284 | 0.0386 | 0.0010 | 0.1428 |

A parallel procedure is applied to the *OrganizationalBarriers* node, where weakly informative Beta priors encode both expected causal strength and uncertainty for each parent. *AnalysisParalysis* is assigned Beta(7.2, 4.8) (mean ≈ 0.60, ESS = 12), and *NoHiring* is assigned Beta(6, 5) (mean ≈ 0.55, ESS = 10), reflecting the view, supported by the literature, that over analysis due to limited ZT knowledge and the lack of qualified security talent are major drivers of organizational fragility during ZTA adoption. *ResistanceToChange* is modeled with Beta(5.5, 6.5) (mean ≈ 0.4583, ESS = 12), indicating a moderate-to-strong effect consistent with empirical findings that staff reluctance to comply with new security policies frequently undermines digital and cybersecurity initiatives. *LegacySystems* is given a slightly lower expected strength via Beta(4, 6) (mean ≈ 0.40, ESS = 10), acknowledging that legacy IT inhibits migration but is often overshadowed by financial and knowledge issues in SMBs. The *AnalysisParalysis* node itself is driven by *Cybersecurityawareness* and *ZTTechKnowledge*, with causal link priors Beta(3.5, 6.5) (mean ≈ 0.35, ESS = 10) and Beta(2.5, 7.5) (mean ≈ 0.25, ESS = 10), respectively. These values encode the belief that low awareness and limited technical expertise around a relatively novel paradigm like ZTA increase confusion and indecision, with cyber awareness exerting a slightly stronger influence. A leak prior distribution of Beta(2, 23) (mean 0.08) is used for *OrganizationalBarriers*, larger than that for FinancialBarriers, to reflect a non-trivial probability of additional unmodeled organizational obstacles (e.g., new regulations, leadership failures, structural issues).

Similarly, MC simulations are implemented in Python, and posterior medians, rather than means, are used as point estimates in GeNIE because the median is less affected by long-tailed distributions and therefore provides a more robust and stable measure of central tendency across repeated runs. The posterior distributions of all causal strengths are summarized by their means, standard deviations, and 95% credible intervals, with medians recorded as the input values (Table 7).

Table 7 The result of MC simulations of the Organizational Barriers node.

| Variable Name | Mean | Median | SD | $q_{2.5\%}$ | $q_{97.5\%}$ |
|---|---|---|---|---|---|
| CybersecurityAwareness | 0.3490 | 0.3388 | 0.14313 | 0.1016 | 0.6469 |
| ZTTechKnow | 0.2514 | 0.2334 | 0.1320 | 0.0493 | 0.5468 |
| Resistancetochange | 0.4571 | 0.4554 | 0.1378 | 0.1982 | 0.7315 |
| LegacySystemSub | 0.3989 | 0.3902 | 0.1474 | 0.1371 | 0.6970 |
| AnalysisParalysis | 0.5988 | 0.6048 | 0.1359 | 0.3267 | 0.8444 |
| NoHiring | 0.6003 | 0.6074 | 0.1478 | 0.3001 | 0.8621 |
| Leak | 0.0802 | 0.0692 | 0.0536 | 0.0099 | 0.2130 |

ZT security measures that act as parent nodes to the ZT pillar nodes are modeled as binary variables, indicating whether a given control is implemented in an SMB or not (Yes/No). This representation reflects the practical, policy-oriented nature of security deployment decisions, which are typically assessed in terms of presence or absence rather than on a continuous scale. The semantics of these parent nodes already encode a positive causal effect (for example, "MFA = Yes" activates the Identity pillar). Accordingly, all ZT security controls are specified as dichotomous variables, with "Yes" denoting that the control is in place and "No" indicating its absence, and their prior probabilities are those previously derived and reported in Table 5.

Because robust empirical estimates of causal strengths for ZT control activation in SMBs are not available, weakly informative Beta priors are assigned using domain knowledge. The prior mean for each strength reflects the expected effect size of a control when present. Core controls like MFA, RBAC, and EDR receive higher prior means than supporting controls such as device inventory (D_INV), consistent with cybersecurity frameworks that prioritize identity and endpoint measures (e.g., NIST CSF). Conversely, controls that are less widely adopted or prone to misconfiguration (e.g., DLP and WAF) are assigned lower prior means, capturing practical deployment challenges. The resulting Beta priors are listed in Table 8, and their posterior strength distributions, summarized by mean, standard deviation, and 95% credible intervals, with posterior medians used as point estimates in GeNIE, are reported in Table 9. The obtained results are validated by computing the prior predictive summary of ZT pillars as illustrated in Appendix A, Table A.1.

MC simulations are run to obtain the mean, median, standard deviation, and 95% credible intervals for the strength parameters of each ZT control. Weakly informative priors are specified at the pillar level: the Data and Device pillars use Beta(8, 8) and Beta(10, 10), respectively (both with mean 0.50); the Identity pillar uses Beta(12, 8) (mean 0.60); and the Application pillar uses Beta(6, 10) (mean 0.38). Table 10 reports the posterior summaries, and the posterior medians are used in GeNIE as the strength values for each pillar. For example, the posterior median for the SSO node represents the probability that SSO alone, in the absence of other Identity controls, is sufficient to make the Identity pillar effective (e.g., by reducing credential risk and enforcing unified authentication).

To compute the aggregated marginal probabilities of the higher-level child nodes *(FinancialBarriers, OrganizationalBarriers, and ZTSMBsPillars)*, uncertainty is propagated via MC sampling using the posterior Beta summaries and link strengths. Each parent marginal (mean and standard deviation) is converted to a Beta distribution via the method of moments, while link strengths are modeled with weakly informative priors. For each of 20,000 draws, parent marginals and link strengths are sampled and combined through the *NoisyOR* function to obtain one draw of the child marginal; empirical means, medians, and 95% credible intervals are then reported. The posterior medians derived from these simulations are also used as strength values in GeNIE for consistency across the model.

Table 8 Causal link prior distributions of ZT control links with pillar nodes.

| ZT Pillars Child Nodes | ZT Security Measures Parent Nodes | Assigned Beta Distribution | Mean |
|---|---|---|---|
| Data | DE | Beta(25, 10) | 0.71 |

| | DLP | Beta(15, 20) | 0.43 |
|---|---|---|---|
| Identity | SSO | Beta(12, 18) | 0.40 |
| | RBAC | Beta(20, 10) | 0.67 |
| | MFA | Beta(28, 8) | 0.78 |
| | DM | Beta(15, 20) | 0.43 |
| Device | D_INV | Beta(18, 12) | 0.60 |
| | EDR | Beta(22, 8) | 0.73 |
| Application | WAF | Beta(15, 25) | 0.38 |

Table 9 A summary of MC simulations of ZT measures-pillars causal links.

| Variable Name | Mean | Median | SD | $q_{2.5\%}$ | $q_{97.5\%}$ |
|---|---|---|---|---|---|
| Data\|DE | 0.713095 | 0.717387 | 0.075121 | 0.556885 | 0.848296 |
| Data\|DLP | 0.428571 | 0.426782 | 0.08303 | 0.27081 | 0.594384 |
| Identity\|SSO | 0.399759 | 0.397125 | 0.088033 | 0.2362 | 0.578668 |
| Identity\|RBAC | 0.666936 | 0.671908 | 0.084354 | 0.489191 | 0.817678 |
| Identity\|MFA | 0.77802 | 0.783767 | 0.068887 | 0.628787 | 0.895944 |
| Device\|DM | 0.427464 | 0.426138 | 0.082209 | 0.271233 | 0.590964 |
| Device\|D_INV | 0.600108 | 0.601829 | 0.087665 | 0.423189 | 0.765604 |
| Device\|EDR | 0.732992 | 0.738279 | 0.080235 | 0.562289 | 0.87307 |
| Application\|WAF | 0.375076 | 0.372517 | 0.075894 | 0.234438 | 0.528217 |

Table 10 A posterior predictive MC summary distributions of ZT pillars' nodes.

| Variable Name | Mean | Median | SD | $q_{2.5\%}$ | $q_{97.5\%}$ |
|---|---|---|---|---|---|
| Data | 0.61525 | 0.6166 | 0.06545 | 0.48323 | 0.73954 |
| Identity | 0.82310 | 0.82566 | 0.04405 | 0.72828 | 0.90116 |
| Device | 0.76767 | 0.77043 | 0.05235 | 0.65766 | 0.86232 |
| Application | 0.17504 | 0.17114 | 0.04527 | 0.09679 | 0.27362 |

The same MC–based procedure is applied to the final child node, *ZTImplementationSuccessChance,* which has three parents: *FinancialBarriers, OrganizationalBarriers*, and *ZTSMBsPillars*. As with earlier parts of the model, the causal link strengths for these parents are given weakly informative Beta priors to reflect domain knowledge while preserving substantial uncertainty. The two barrier nodes are modeled as *inhibitors* so that, semantically, their activation reduces the probability of successful ZT implementation, whereas *ZTSMBsPillars* functions as an enabler. For each barrier, a sampled causal strength represents the probability that, when the barrier is present, it prevents the child node from activating.

Parent-node uncertainties are represented with Beta priors, and the causal strengths are likewise treated as random variables with weakly informative Beta distributions. In each MC iteration, samples for parent marginals and link strengths are drawn and combined via the *NoisyOR* formula to compute the activation probability of the child node, thereby propagating epistemic uncertainty through the Bayesian Network structure.

As reported in Table 11, the causal link from ZTSMBsPillars is assigned a Beta(12, 8) prior (mean 0.6, ESS = 20), reflecting its role as a relatively strong positive contributor to successful ZT implementation in SMBs. *FinancialBarriers* and *OrganizationalBarriers* are given Beta(6.4, 9.6) and Beta(7.2, 8.8) priors (means 0.4 and 0.45, respectively), representing moderate inhibitory effects consistent with their constraining influence in practice. Table 12 presents the posterior summaries for these three parent nodes, showing mean and median values that confirm a moderate-to-strong net effect: the *ZTSMBsPillars* node acts as a strong enabler, while the barrier nodes exert noticeable but not overwhelming negative impacts. The relatively large standard deviations and wide 95% credible intervals indicate substantial epistemic uncertainty about the exact strength of each causal link, an expected and desirable property when working with weakly informative priors and limited hard evidence, as it guards against overconfidence in the model. For implementation in GeNIE, the posterior medians of the causal strengths are used as point estimates when specifying the NoisyOR links, ensuring robust and stable parameterization of the *ZTImplementation-SuccessChance* node.

Table 11 Causal link prior distributions of parent nodes of the final child nodes.

| ZT Pillars Child Nodes | ZT Security Measures Parent Nodes | Assigned Beta Distribution | Mean |
|---|---|---|---|
| *ZTImplementationSuccessChance* | *ZTSMBsPillars* | Beta(12, 8) | 0.60 |
| | *OrganizationalBarriers* | Beta(7.2, 8.8) | 0.45 |
| | *FinancialBarriers* | Beta(6.4, 9.6) | 0.40 |

Table 12 A summary of MC simulations of the Implementation Success Parent nodes' causal links.

| Variable Name | Mean | Median | SD | $q_{2.5\%}$ | $q_{97.5\%}$ |
|---|---|---|---|---|---|
| *ZTSMBsPillars* | 0.5999 | 0.6039 | 0.1074 | 0.3824 | 0.7980 |
| FinancialBarriers | 0.4005 | 0.3960 | 0.1183 | 0.1815 | 0.6394 |
| OrganizationalBarriers | 0.4497 | 0.4476 | 0.1211 | 0.2217 | 0.6904 |

The Implementation Success Model (ISM) was implemented and evaluated using the GeNIE platform, with the resulting graphical structure displayed in Figure 3. After Bayesian inference, the marginal posterior probability of the final child node, ZTImplementationSuccessChance, is 0.6510, indicating a 65.10% chance of successfully implementing ZTA in SMBs under the modeled conditions and with the included financial and organizational barriers. This value is obtained by entering the point estimates (posterior medians and leak parameters) derived from the MC simulations into GeNIE.

Figure 5 further shows that, given the specified priors and causal strengths informed by domain knowledge, the posterior probabilities that *FinancialBarriers* and *OrganizationalBarriers* are active are approximately 0.61 and 0.63, respectively. This implies that, on average, there is a 61% likelihood of financial hurdles and a 63% likelihood of organizational obstacles arising during ZTA implementation for SMBs in the modeled scenario.

At the same time, the ZTSMBsPillars node has an estimated posterior probability of 0.68, indicating that the modeled ZT security measures and pillars together provide a 68% maturity or readiness level. In practical terms, this suggests that while the technical Zero Trust controls can reach a relatively strong readiness state, their positive effect on ZT implementation success is partially offset by the substantial probability of concurrent financial and organizational barriers.

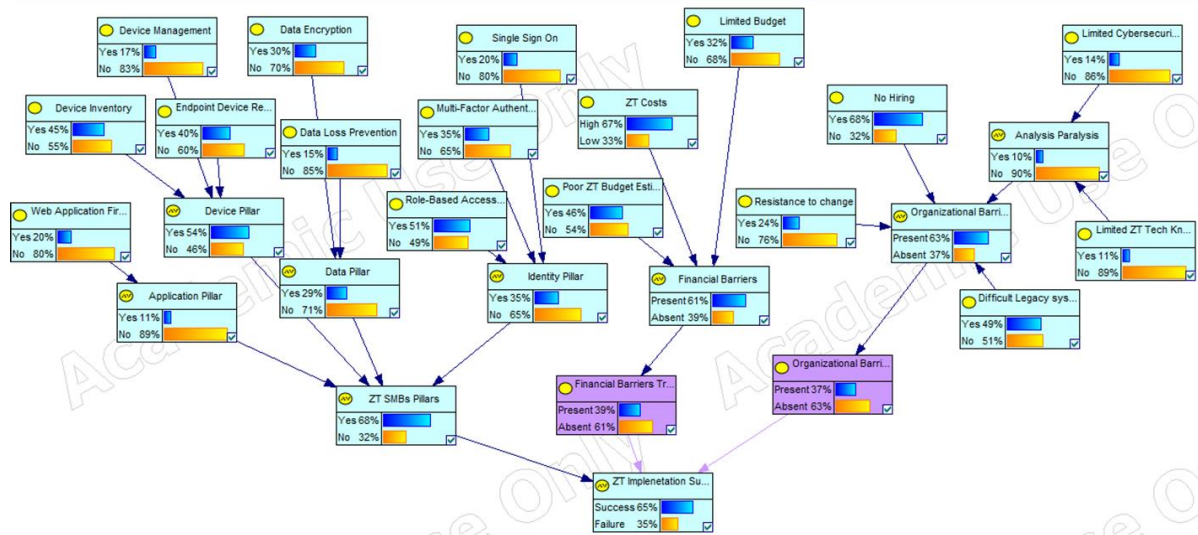

Figure 5 The BN of ZT implementation success sub-model.

## 3.5 Quantifying BN-RAM Parameters

3.5.1 Calculations of the Prior Probabilities of RAM's Parent Nodes

The prior probabilities for the four attack vectors in the BN-RAM are estimated using a frequency-based approach applied to those Advisen incidents that specifically involve U.S. SMBs. A Python-based algorithm first prepares the data by importing the subset returned by Algorithm 1 into a DataFrame, then standardizes column names (e.g., removing spaces and converting to uppercase) to support consistent querying. The core of the method is a rule-based classification scheme where each attack type is associated with criteria spanning three fields: case type, case description, and actor type, so that incidents are classified based on multiple evidence points rather than a single attribute.

For Insider Threat, incidents are directly labeled when the actor type is "internal"; if the actor field is missing, the algorithm inspects case type and description to identify qualifying events, thus reducing false positives and improving accuracy. Phishing and Physically Lost or Stolen Asset cases are detected by cross-checking case type and description for relevant terms, with actor type used as an additional discriminator. Credential-Based Attacks are identified by matching case type, description, or actor type against a predefined keyword list, and the full pseudo-code for the logic of Algorithm 2 is shown in Figure A.2, in Appendix A. Boolean masks are then applied to filter incidents per attack type, and the row counts of these subsets yield empirical frequencies, which form the basis for prior probability estimation.

Finally, prior probabilities are computed using Laplace smoothing, as specified in Equation 22, by adding 1 to the count of incidents for each attack type and K (the number of attack categories) to the denominator. This adjustment prevents any attack vector from being assigned a zero probability when no incidents are observed in the filtered dataset, ensuring numerically stable priors for subsequent Bayesian modeling.

$$P(attack) = \frac{Attack\ count + 1}{total\ instances + K} \quad \text{(Equation 22)}$$

Where:
Attack count = number of cases matching the filtering rules,
Total instances = total number of cases in the dataset,
K = number of the attack vectors considered.

Running the classification algorithm on the Advisen subset yields a set of incidents mapped to each of the four attack categories, enabling prior probability estimation for every attack node in the BN-RAM. Incidents that satisfy the predefined, rule-based criteria are counted per attack type, and these counts are converted into probabilities via Equation 22 with Laplace smoothing, which avoids zero probabilities for rare or unobserved events. The resulting normalized priors capture the empirical distribution of cyber incidents affecting U.S. SMBs and provide the basis for subsequent probabilistic inference in the Bayesian model.

To validate these results, incidents associated with each attack vector were repeatedly exported to separate CSV files and manually checked for issues such as double-counting or misclassification, with any anomalies corrected to enhance confidence in the final probabilities. The algorithm executed efficiently, completed in 0.524 seconds on a Dell laptop (Intel(R) Core(TM) i7-9850H, 16 GB RAM) using a cloud-based Python environment.

Empirically, 48 incidents were classified as Phishing, yielding a prior probability of 3.3%, indicating that, while present in the dataset, phishing is less frequent than other modeled attacks. Credential-Based attacks accounted for 510 incidents (≈34%), suggesting that misuse or theft of credentials is a common mechanism in reported cases. Insider Threats were the most prevalent, with 713 incidents and a prior probability of about 48%, underscoring the prominence of internal, intentional or accidental actors in the dataset. Physically Lost or Stolen Assets ranked third, with 96 incidents and an 11% prior probability. An additional 59 of the 1,486 total incidents did not fit any of the four categories and thus remained unassigned, as they correspond to other attack types outside the model's current scope. Table 13 summarizes these priors, where each attack-node is modeled with two states (True/False), and the "True" state probability equals the estimated prior for that attack vector.

Table 13 The prior probabilities for each attack type.

| Attack Type | Modeling Technique | Incident Count | Prior Probability Value |
|---|---|---|---|
| Phishing | Discrete | 48 | 3.29% |
| Credential-Based Attacks | Discrete | 510 | 34.30% |
| Insider Threats | Discrete | 713 | 47.92% |
| Physically Lost or Stolen Assets | Discrete | 156 | 10.54% |

3.5.2 Calculations of the Conditional Probabilities for Assets Nodes

The conditional probability tables (CPTs) for each asset node in the BN_RAM are derived by feeding the outputs of Algorithm 2 into Algorithm 3, which estimates the probability of a given asset being compromised conditional on a specific attack type, as in Equations 23. Algorithm 3 operates on the "PROXIMATE_CAUSE" field in the Advisen dataset, which identifies the primary asset affected in each incident. These textual entries are first normalized via a mapping scheme into standardized asset categories (Server, Email, Website, Printed Records, User Devices, Software, Portable Data Storage, etc.). For each attack type, the algorithm then counts how often each asset category appears and converts these counts into conditional probabilities by dividing by the total number of incidents assigned to that attack, thereby producing CPTs that capture $P(Asset \mid Attack\ Type)$.

Where:
$Asset \in \{Server, Email, Website, Laptop/Desktop, \ldots\}$
$Attack \in \{Phishing, Credential\text{-}Based\ Attacks, Insider\ Threats, Physically\ Lost/Stolen\}$
For a given attack type $a$ and an asset type $x$:

$$P(Asset = x \mid Attack = a)$$
$$= \frac{Number\ of\ incidents\ where(Asset = x \mid Attack = a)}{Total\ number\ of\ incidents\ where\ Attack = a} \quad \text{(Equation 23)}$$

To validate Algorithm 3, several internal consistency checks are applied. The number of missing "PROXIMATE_CAUSE" values and the distribution of incidents across asset categories are compared against the algorithm's outputs, and the total number of rows in the full assignment table from Algorithm 2 (1,486) is compared with the 1,427 processed records. The difference is attributed to blank proximate-cause entries and incidents involving asset types outside the model's scope. This verification confirms that the remaining incidents are correctly mapped and that the derived CPTs are internally consistent; a pseudo-code representation of Algorithm 3 is shown in Appendix A, Figure A.3.

The resulting conditional probability distributions highlight how different attack vectors concentrate on distinct assets. For phishing, 93.5% of mapped incidents involve Email as the impacted asset, reflecting phishing's reliance on electronic communications, while 6.5% involve Servers, likely representing downstream compromise following initial email-based intrusion. Credential-Based attacks predominantly target Servers (84.5%), with smaller probabilities for Websites (7%) and Emails (6%), indicating that stolen credentials are often used to access high-value back-end systems but can also be leveraged against front-end and messaging systems; low-frequency links to User Devices, Printed Records, and Portable Storage show broader, though rarer, credential misuse.

Insider Threats exhibit a markedly different pattern. Printed Records account for 57.9% of cases, underscoring insiders' unique ability to abuse access to physical documents containing sensitive data, while Servers (27.3%) and Emails (7.2%) are also common targets; Websites, Portable Storage, and Software appear with lower probabilities, illustrating the hybrid physical–digital nature of insider risk. For Physically Lost or Stolen Assets, User Devices (laptops, desktops) dominate at 62.5%, followed by Printed Records (28.9%) and Portable Storage (7.8%), with a small fraction involving Servers (0.8%), consistent with the predominantly physical modality of this vector. These conditional probabilities are summarized in Table 14.

Within GeNIE, each attack node is modeled as a binary variable (True/False) indicating whether the attack occurs. Asset nodes are also binary (True = asset compromised, False = not compromised) and use the *NoisyOR* function so that the conditional probabilities estimated above inform how the occurrence of each attack type probabilistically activates compromise of the corresponding asset.

Table 14 The values of conditional probabilities for each asset given the occurrence of a certain attack type.

| Attack Type | Targeted Asset(s) | Incident Count | Conditional Probability Value |
|---|---|---|---|
| Phishing | Email | 43 | 0.9348 |
|  | Server | 3 | 0.0652 |
| Credential-Based Attacks | Server | 267 | 0.8449 |
|  | Website | 22 | 0.06962 |
|  | Email | 19 | 0.060127 |
|  | User Devices | 5 | 0.015823 |
|  | Printed Records | 2 | 0.006329 |
|  | Portable Data Storage Instruments | 1 | 0.003165 |
| Insider Threats | Printed Records | 291 | 0.579681 |
|  | Server | 137 | 0.272908 |
|  | Email | 36 | 0.071713 |
|  | Website | 20 | 0.039841 |
|  | Portable Data Storage Instruments | 11 | 0.021912 |
|  | User Devices | 5 | 0.00996 |
|  | Software | 2 | 0.003984 |
| Physically Lost or Stolen Assets | User Devices | 80 | 0.625 |
|  | Printed Records | 37 | 0.289063 |
|  | Portable Data Storage Instruments | 10 | 0.078125 |
|  | Server | 1 | 0.007813 |

3.5.3 Calculations of Causal Link Strength Values between Intermediate Cause Nodes and Asset Nodes

Intermediate nodes labeled *Cause_Attack_Asset* are introduced to operationalize the inhibitory effect of ZT control (ZTC) nodes and to quantify how controls reduce the likelihood that a given attack successfully compromises a specific asset. These intermediate nodes sit between attack nodes, ZTC nodes, and asset nodes: they are children of both attack and ZTC nodes, and parents of the asset nodes, and are modeled using the *NoisyOR* function. In this structure, causal links from ZTCs to the intermediate nodes encode the probability that a given control mitigates or blocks an attack on an asset, while links from attacks to the intermediate nodes represent the conditional probability that an attack, when it occurs, targets that asset. The links from *Cause_Attack_Asset* nodes to asset nodes then capture the probability that, if the combined "attack on asset" event succeeds (after accounting for ZTCs), the asset actually becomes compromised.

The conditional inputs for these intermediate nodes are derived from two sources: (i) empirical attack–asset probabilities estimated from historical data (Table 14), and (ii) causal strengths of ZTCs obtained from prior modeling in the ISM. Monte Carlo (MC) simulation is used to compute posterior summaries for all links, with Beta priors for each causal strength informed by literature and industry reports. First, MC simulations are run to propagate the posterior Beta distributions of ZT security measures from the ISM into their corresponding ZTC nodes in the RAM, using the same strength-estimation approach as in the ISM. The resulting posterior summaries for each ZTC node are reported in Table 15, and posterior medians are used as point-strength values when integrating the ISM and RAM in GeNIE.

Table 15 The MC summary of the posterior distributions for each ZTC node.

| Variable Name | Mean | Median | SD | $q_{2.5\%}$ | $q_{97.5\%}$ |
|---|---|---|---|---|---|
| ZTC1 | 0.404867 | 0.403433 | 0.065128 | 0.281647 | 0.534357 |
| ZTC3 | 0.748874 | 0.754354 | 0.075621 | 0.588829 | 0.880697 |
| ZTC2 | 0.726183 | 0.728345 | 0.050216 | 0.622677 | 0.817859 |
| ZTC4 | 0.199634 | 0.195665 | 0.055912 | 0.101851 | 0.319107 |

A Python code is developed and utilized to estimate the causal strengths between ZTC nodes and intermediate nodes through running MC simulations. Weakly or moderately informative Beta priors for these links are set using evidence from sources such as Verizon DBIR [59], Zscaler [146], NIST SP 800-207 [147], and National Security Agency ZT guidance [124, 128, 148], as detailed in Table A.2 (Appendix A). For each of 20,000 MC iterations, strength samples are drawn from these Beta priors, combined via the *NoisyOR* formula to represent the probability that at least one ZTC sufficiently mitigates the risk, and then merged with attack–asset priors and a leak prior of Beta(2, 48). This process yields posterior distributions for each ZTC–intermediate link, summarized in Tables A.3–A.6 in Appendix A. The same approach is applied to estimate the causal strengths from intermediate nodes to asset nodes, where each strength represents the probability that, given an attack has succeeded despite ZTCs, a specific asset is actually compromised.

In practice, priors for these intermediate–asset links are set close to 1, reflecting that once an attack bypasses all relevant ZTCs, asset compromise is almost certain. Exceptions are made where additional, independent safeguards exist (e.g., physical controls or human checks for Printed Records), and lower prior means are used; accordingly, these priors are listed in Table A.3 (Appendix A). Posterior samples for intermediate nodes (parents) are then fitted to Beta distributions via the method of moments (matching mean and standard deviation), and these fitted Betas are used in subsequent MC runs to propagate uncertainty through the network. The fitted parameters (e.g., $\alpha, \beta$) for the intermediate-node posteriors are reported in Table A.7, Appendix A, and the final MC posterior predictive summaries for the intermediate–asset causal links are presented in Table 16.

Table 16 The posterior predictive summary of the causal links between Intermediate nodes and asset nodes.

| Attack | Asset | Mean | Median | SD | $q_{2.5\%}$ | $q_{97.5\%}$ |
|---|---|---|---|---|---|---|
| Phishing | Email | 0.9008 | 0.9267 | 0.0905 | 0.6639 | 0.9972 |
| Insider Threat | Email | 0.7008 | 0.7147 | 0.1377 | 0.4016 | 0.9275 |
| Credential-Based | Email | 0.9008 | 0.9265 | 0.0896 | 0.6670 | 0.9973 |
| Credential-Based | Server | 0.8997 | 0.9259 | 0.0910 | 0.6624 | 0.9972 |
| Insider Threat | Server | 0.6995 | 0.7119 | 0.1381 | 0.3983 | 0.9252 |
| Phishing | Server | 0.8007 | 0.8198 | 0.1188 | 0.5234 | 0.9717 |
| Physically Lost or Stolen | Server | 0.6017 | 0.6099 | 0.1478 | 0.3023 | 0.8640 |
| Credential-Based | Website | 0.8011 | 0.8215 | 0.1201 | 0.5208 | 0.9717 |
| Insider Threat | Website | 0.6998 | 0.7151 | 0.1387 | 0.4031 | 0.9257 |
| Credential-Based | Printed Records | 0.6000 | 0.6061 | 0.1475 | 0.3027 | 0.8643 |
| Insider Threat | Printed Records | 0.4992 | 0.5002 | 0.1515 | 0.2108 | 0.7916 |
| Physically Lost or Stolen | Printed Records | 0.7002 | 0.7132 | 0.1379 | 0.3997 | 0.9255 |
| Insider Threat | Software | 0.6001 | 0.6071 | 0.1474 | 0.3016 | 0.8640 |
| Credential-Based | User Devices | 0.7991 | 0.8200 | 0.1210 | 0.5193 | 0.9718 |
| Physically Lost or Stolen | User Devices | 0.7991 | 0.8203 | 0.1210 | 0.5164 | 0.9699 |
| Insider Threat | User Devices | 0.6993 | 0.7132 | 0.1376 | 0.4022 | 0.9249 |

| Attack | Asset | Mean | Median | SD | $q_{2.5\%}$ | $q_{97.5\%}$ |
|---|---|---|---|---|---|---|
| Credential-Based | Portable Data Storage Instruments | 0.8006 | 0.8211 | 0.1200 | 0.5226 | 0.9720 |
| Physically Lost or Stolen | Portable Data Storage Instruments | 0.8000 | 0.8194 | 0.1198 | 0.5175 | 0.9718 |
| Insider Threat | Portable Data Storage Instruments | 0.6984 | 0.7122 | 0.1386 | 0.3999 | 0.9242 |

3.5.4 Data Breach Node Computations

To evaluate which attack vector is most likely to result in a data breach in the presence of ZT security measures (RQ3), a Data Breach node is introduced as a common child of all asset nodes in the BN_RAM. This node is positioned downstream of the assets and represents the event that compromise of a given asset leads to a breach of sensitive information. Its inputs are asset-specific strength values, each reflecting the probability that compromise of that asset alone (in the absence of other contributing assets) results in a data breach. Because the mitigating effect of ZT controls (ZTCs) has already been modeled at the asset layer, the Data Breach node depends only on asset-compromise states and does not directly include ZTCs as parents, thereby avoiding double-counting their influence. The node is modeled via a *NoisyOR* function parameterized by asset-to-breach strengths, which are denoted by Equation 24, where $s_j$ is the strength of the asset $j$.

$$s_j = P(Data\ Breach\ | Asset_j = 1, All\ other\ assets = 0) \quad \text{(Equation 24)}$$

These strength parameters are estimated empirically from the SMB incident subset produced by Algorithm 2. An additional "leak" parameter is also derived from the same data to capture breaches originating from assets outside the modeled set. For this purpose, a dedicated algorithm filters the 1,427 incidents (associated with the four modeled attack types) to identify data-breach events and map them to assets. First, incidents labeled with case type "Malicious Breach" and those whose case description contains the keyword "breach" are selected. Each breach incident is then mapped to an asset using a keyword-based function on the PROXIMATE_CAUSE field, with de-duplication enforced via the unique MSCAD_ID identifier. Incidents that cannot be mapped to any of the modeled assets (because of blank proximate-cause entries or different asset types) are used to estimate the leak term; 84 such cases are identified as unmapped.

The probability of breaches $P_j$ involving the asset $j$ is provided in Equation 25, where $n_j$ is the number of breach incidents linked to asset $j$ and $N$ is the total number of breach events.

$$P_j = \frac{n_j}{N} \quad \text{(Equation 25)}$$

The leak probability is then calculated via Equation 26 as the proportion of breaches not mapped to any modeled asset.

$$l = \frac{n_{unmapped}}{N} \quad \text{(Equation 26)}$$

Finally, the asset strength $s_j$ is determined using Equation 27, representing the conditional probability that the compromise of the asset $j$ propagates to a data breach, while the leak term captures residual risk from unmodeled or unknown assets. The resulting strength values and leak estimate, summarized in Table 17, are used as the *NoisyOR* parameters for the *DataBreach* node in the BN, enabling quantification of how different attack vectors, via their impact on assets, contribute to breach risk under ZT controls.

$$s_j = \frac{P_j}{1 - l} \qquad \text{(Equation 33)}$$

Table 17 Results of Algorithm 4.

| Asset | Count | $P_j$ | $s_j$ |
|---|---|---|---|
| Server | 355 | 0.538695 | 0.6174 |
| Email | 55 | 0.08346 | 0.0957 |
| Website | 28 | 0.042489 | 0.0487 |
| Printed Records | 102 | 0.15478 | 0.1774 |
| User Devices | 25 | 0.037936 | 0.0435 |
| Portable Data Storage Instruments | 10 | 0.015175 | 0.0174 |

3.5.5 The Computations of the Integrated Model

The integrated model connects the Implementation Success Model (ISM) and the Risk Analysis Model (RAM) in two main ways. First, the ZT security measures from the ISM act as parents to, and collectively define, the ZT Control (ZTC) nodes used in the RAM. Second, the final ISM outcome node, ZTImplementationSuccessChance, is linked directly to the overall response variable, *RiskMagnitude*, which is itself a child of the *DataBreach* node in the RAM.

As in the two sub-models, MC simulations are used to estimate the strengths of the causal links from these two high-level parents, *DataBreach* and *ZTImplementation-SuccessChance*, to the *RiskMagnitude* node. On the RAM side, the marginal probabilities of asset compromise, the asset-to-breach strengths, and the leak parameter (from Table 17) are combined via the *NoisyOR* formula to propagate uncertainty from assets to the *DataBreach* node, and then from *DataBreach* to *RiskMagnitude*. A Beta(16, 4) prior (mean 0.8) is assigned to the *DataBreach* → *RiskMagnitude* link to represent a strong prior belief that breaches substantially increase risk, with moderate confidence. In each MC iteration, asset strengths and breach probabilities are sampled and aggregated using *NoisyOR*, after which a strength sample for *DataBreach* → *RiskMagnitude* is drawn from its Beta prior, and posterior predictive distributions are obtained.

On the ISM side, the posterior mean and standard deviation of *ZTImplementation-SuccessChance* are used to fit a Beta distribution for this node, which is then used in MC simulations. The causal link *ZTImplementationSuccessChance* → *RiskMagnitude* is given a Beta(17, 7) prior (mean 0.71), reflecting moderate confidence, based on domain knowledge and industry reports such as the Okta report, that higher ZT implementation success reduces overall cyber risk, provided financial and organizational barriers are sufficiently mitigated. MC simulations sample from both link priors to generate posterior predictive distributions for the two parent nodes after integrating ISM and RAM, summarized in Table 18.

Table 18 The posterior predictive summary of the parent nodes of the response variable.

| Variable Name | Mean | Median | SD | $q_{2.5\%}$ | $q_{97.5\%}$ |
|---|---|---|---|---|---|
| *ZTImplementationSuccess* | 0.708062 | 0.714176 | 0.091865 | 0.51338 | 0.869053 |
| *DataBreach* | 0.79945 | 0.809223 | 0.087963 | 0.603552 | 0.939363 |

# 4. Results and Analysis

## 4.1 Results of the Base-Case Scenario of the Integrated Model

In the integrated base scenario, the full Bayesian model is executed in GeNIE to connect the two sub-models and visualize their interactions. As illustrated in Figure 6, the ZTCs are linked upward to their corresponding ZT security measures, while the *ZTImplementationSuccessChance* node and the *DataBreach* node both feed into the final response node, *RiskMagnitude*. For interpretability, nodes are color-coded: asset nodes appear in grey, attack vectors in red, ZT mitigation controls in green at the center, the *DataBreach* node in aqua, and the *RiskMagnitude* node in orange. Purple nodes serve as dummy structures to operationalize inhibitory effects, for example, the financial and organizational barrier nodes are modeled so their negative influence on implementation success is reversed as uncertainty propagates through these dummy nodes to the child node.

Each node is annotated with a probability bar and a numeric percentage representing the relevant prior or posterior probability, given the propagated uncertainty and base-case assumptions. Under the baseline configuration, reflecting the current maturity of the ZTCs, Bayesian inference yields a posterior probability of 71% that a data breach occurs, and a corresponding posterior risk magnitude of 63%.

## 4.2 Forward and Backward Analyses

Forward and backward propagation analyses are applied to explore how uncertainty influences the integrated Bayesian model and to support decision makers in formulating strategies to manage that uncertainty. These analyses also allow users to pose queries at any node and observe how beliefs update throughout the network.

Forward propagation follows a cause-and-effect logic, where changes in the marginal distributions of parent nodes are used to quantify their impact on connected child nodes. In contrast, backward propagation begins by fixing evidence on a target (successor) node and re-running inference so that updated probabilities are propagated in reverse, adjusting the beliefs about upstream (predecessor) nodes across the entire network [74].

### 4.2.1 Forward Analysis

Forward propagation analysis is used to evaluate how different disruptive scenarios affect ZT implementation success and overall cyber risk in SMBs. In the first scenario, financial stressors are examined by setting both *LimitedBudget* and *ZTCosts* to True, which raises the probability of FinancialBarriers to 91%, lowers the *ZTImplementationSuccessChance* to 57%, and slightly increases the *RiskMagnitude* to 64%, relative to the base case shown in Table 19. Additional scenarios (3–5) explore organizational stressors by activating AnalysisParalysis, NoHiring, ResistanceToChange, and *DifficultLegacySystemSubstitution* under the *OrganizationalBarriers* node; this combination increases the OrganizationalBarriers probability from 63% to 95%, reduces implementation success from 57% to 47%, and raises the risk level to 65%.

Table 19 Forward propagation analysis of ZT Barriers and Its Impact on Implementation Success and Risk.

| Scenario | Limited Budget | ZT Costs | Analysis Paralysis | No Hiring | Resistance to Change | Difficult Legacy System Substitution | Financial Barriers (%) | Organizational Barriers (%) | ZT Implementation Success (%) | Risk (%) |
|---|---|---|---|---|---|---|---|---|---|---|
| Base | – | – | – | – | – | – | 61 | 63 | 65 | 63 |
| 1 | Y | H | – | – | – | – | 91 (↑) | 63 | 57 (↓) | 64 (↑) |
| 2 | Y | H | Y | – | – | – | 91 | 84 (↑) | 49 (↓) | 65 (↑) |
| 3 | Y | H | Y | Y | – | – | 91 | 90 (↑) | 49 | 65 |
| 4 | Y | H | Y | Y | Y | – | 91 | 94 (↑) | 48 (↓) | 65 |
| 5 | Y | H | Y | Y | Y | Y | 91 | 95 (↑) | 47 (↓) | 65 |

↓ *indicates a decrease in the corresponding variable compared to its value in the base scenario*
↑ *indicates an increase in the corresponding variable compared to its value in the base scenario*
*Y refers to the "Yes" state, while H refers to the "High" state*

A second group of forward scenarios examines how improving ZT pillar maturity influences both implementation success and risk. In Scenario 1, enabling all Identity controls (MFA, SSO, RBAC) increases the Identity pillar maturity to 80%, elevates overall ZT maturity to 84%, raises the *ZTImplementationSuccessChance* to 71%, and reduces *RiskMagnitude* to 59%, as reported in Table 20. Scenarios 2–4 show that fully activating the Devices, Data, and Application pillars raises their respective maturities to 94%, 84%, and 39%, driving overall ZT maturity to 96% and increasing implementation success to 75%. Under these conditions, the risk level is reduced by 14 percentage points compared to the base case, indicating that higher ZT maturity in SMBs can substantially mitigate cyber risk when the modeled controls are effectively deployed.

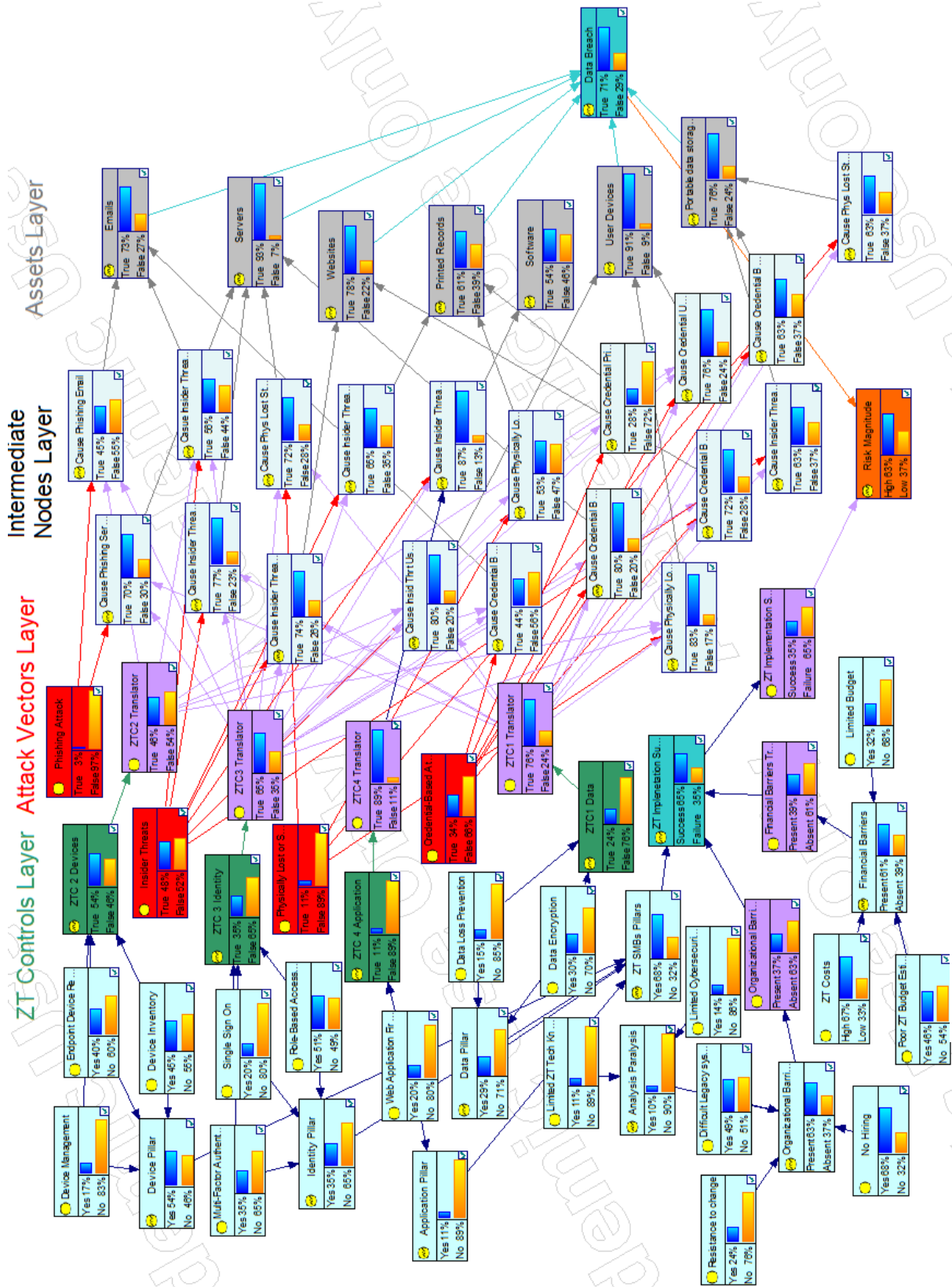

Figure 6 The base BN model after integrating ISM and RAM.

Table 20 Forward propagation analysis of ZT Maturity Level and Its Impact on Risk.

| Scenario | MFA | SSO | RBAC | DM | D_INV | EDR | DE | DLP | WAF | Identity Pillar (%) | Device Pillar (%) | Data Pillar (%) | Application Pillar (%) | ZT SMBs Pillars (%) | ZT Implementation Success (%) | Risk (%) |
|---|---|---|---|---|---|---|---|---|---|---|---|---|---|---|---|---|
| Base | – | – | – | – | – | – | – | – | – | 35 | 54 | 29 | 11 | 68 | 65 | 63 |
| 1 | Y | Y | Y | – | – | – | – | – | – | 80 (↑) | 54 | 29 | 11 | 84 (↑) | 71 (↑) | 59 (↓) |
| 2 | Y | Y | Y | Y | Y | Y | – | – | – | 80 | 94 (↑) | 29 | 11 | 93 (↑) | 74 (↑) | 58 (↓) |
| 3 | Y | Y | Y | Y | Y | Y | Y | Y | – | 80 | 94 | 84 (↑) | 11 | 96 (↑) | 75 (↑) | 49 (↓) |
| 4 | Y | Y | Y | Y | Y | Y | Y | Y | Y | 80 | 94 | 84 | 39 (↑) | 96 | 75 | 49 |

Forward propagation analysis is conducted to demonstrate the effectiveness of ZTCs in reducing the likelihood that modeled assets are compromised by cyberattacks, and to quantify the downstream effects on data breach probability and overall risk magnitude. Four scenarios are simulated, each incrementally activating additional ZTCs, with results summarized in Table 21. The table reports, for each scenario, the conditional probability that individual assets are compromised alongside the resulting data breach likelihood and risk level.

In Scenario 1, implementing ZTC1 moderately lowers the breach probability for servers and websites, reducing the overall risk to 56%. Scenario 2, which includes both ZTC1 and ZTC2, further decreases compromise probabilities for emails and user devices, yielding a risk level of 55%. Scenario 3 adds ZTC3 and produces substantial reductions in compromise likelihood, particularly for emails, servers, websites, and portable data storage, with the risk level dropping significantly to 40%. Finally, Scenario 4 activates all ZTCs (ZTC4), providing strong protection for software while maintaining the data breach probability and risk level achieved in Scenario 3. These findings confirm that a comprehensive, layered ZTA deployment meaningfully reduces both the probability of data breach incidents and SMBs' aggregate cyber risk profile, thereby strengthening their cybersecurity posture.

Table 21 Forward propagation analysis of ZTCs' Effectiveness in reducing the data breach and risk level.

| Scenario | ZTC1 | ZTC2 | ZTC3 | ZTC4 | Emails | Servers | Websites | Printed Records | Software | User Devices | Portable Data Storage | Instruments | Data Breach | Risk (%) |
|---|---|---|---|---|---|---|---|---|---|---|---|---|---|---|
| Base | – | – | – | – | 73 | 93 | 78 | 61 | 54 | 91 | 76 | 71 | 63 | |
| 1 | Y | – | – | – | 73 | 74 (↓) | 46 (↓) | 61 | 54 | 73 (↓) | 13 (↓) | 62 (↓) | 56 (↓) | |
| 2 | Y | Y | – | – | 57 (↓) | 74 | 46 | 61 | 54 | 60 (↓) | 13 | 61 (↓) | 55 (↓) | |
| 3 | Y | Y | Y | – | 18 (↓) | 42 (↓) | 11 (↓) | 23 (↓) | 50 (↓) | 16 (↓) | 13 | 40 (↓) | 40 (↓) | |
| 4 | Y | Y | Y | Y | 18 | 42 | 11 | 23 | 6 (↓) | 16 | 13 | 40 | 40 | |

### 4.2.2 Backward Analysis

Backward propagation analysis updates beliefs across the entire network when evidence is set on a specific target node, with uncertainty flowing upstream to predecessor nodes. In GeNIE, this is implemented via virtual evidence, which introduces uncertain observations as likelihood ratios rather than fixed states, a probability distribution over possible outcomes for unobservable variables. The resulting posterior probabilities integrate this virtual evidence with the network structure and parent priors, such that strong confirming prior evidence can cause the posterior to exceed the assigned virtual evidence probability, reflecting comprehensive belief updating across the model.

Scenario 1 sets virtual evidence on the Risk node at 20% and examines the required changes in upstream nodes. To achieve this low-risk target, ZTImplementationSuccess must increase to 69%, with ZTC1, ZTC2, and ZTC3 maturities rising to 26.6%, 54.6%, and 37.2%, respectively. The DataBreach probability must also fall to 50.9%, demonstrating that substantial ZTC maturity improvements are needed to reach this risk threshold. Table 22 compares marginal posterior probabilities before and after setting virtual evidence, confirming these required adjustments. Scenario 2 sets hard evidence on *DataBreach* = 100% True and applies GeNIE's Most Probable Explanation (MPE) algorithm to identify the most likely configuration of *all* model nodes given this evidence. The MPE algorithm maximizes the posterior probability over state assignments and reports three quantities: *P(MPE|E)*, the posterior probability of the MPE assignment given evidence *E*; *P(E)*, the marginal likelihood of the evidence; and *P*(MPE,*E*), the joint probability of both. Figure 7 displays the most probable state combination across the network nodes under confirmed data breach conditions.

Table 22 The analyzed output of the backward propagation analysis of scenario 1.

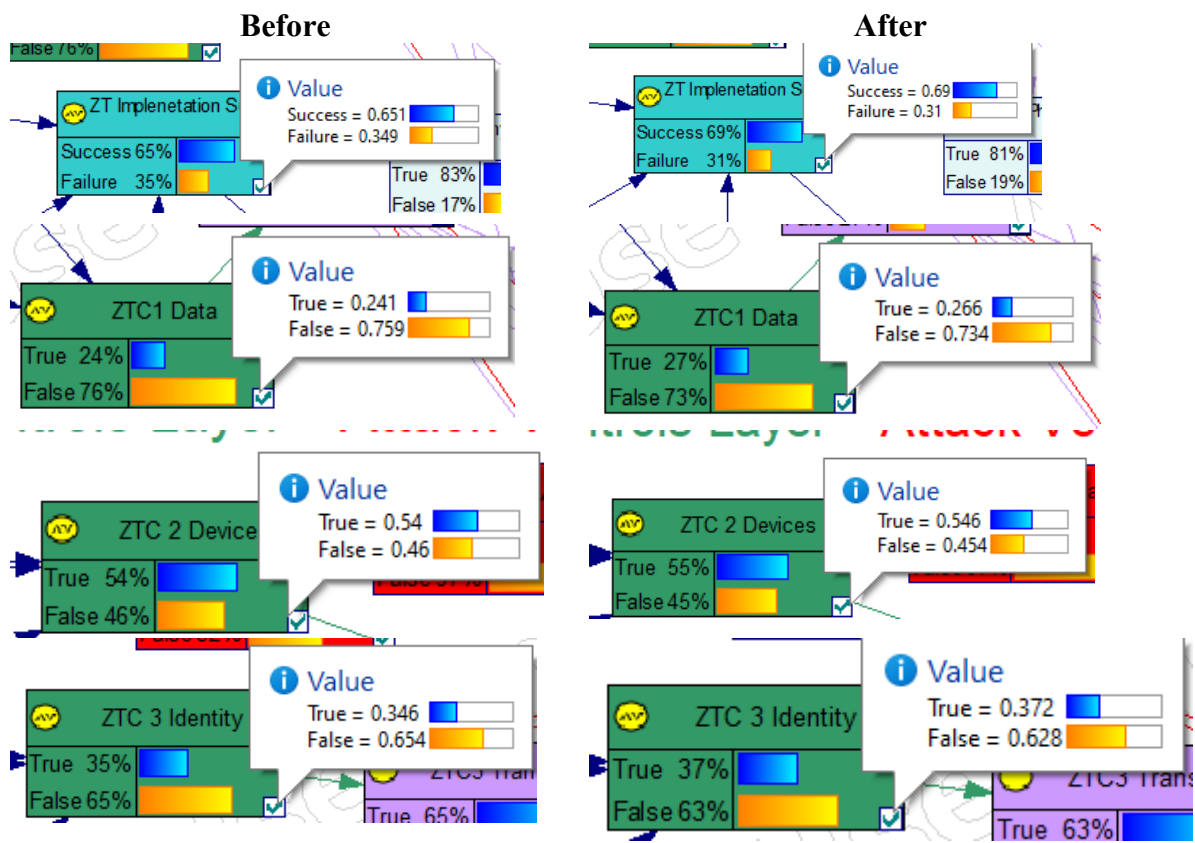

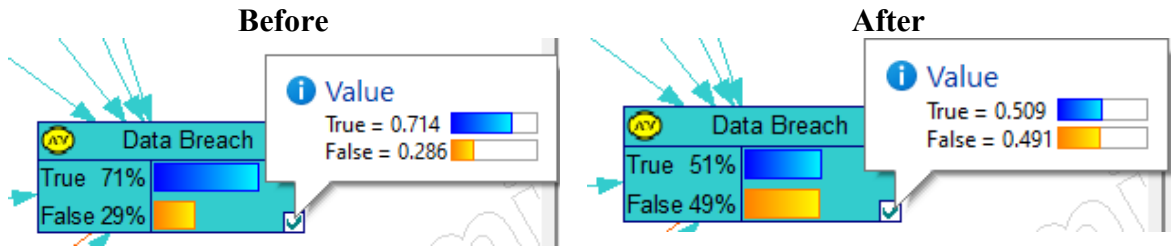

The key findings from Scenario 2's MPE analysis reveal that a confirmed data breach is most likely when ZTC maturity levels are uniformly low across all pillars, multiple attack vectors are active, and most assets are compromised. Financial barriers (specifically ZTCosts = True) and organizational barriers (*DifficultLegacySystemSubstitution* = True and *NoHiring* = True) are also active in this highest-probability configuration. Table 23 reports the marginal posterior probabilities under confirmed breach conditions, showing that Identity and Data ZTC maturities decrease by 4% while risk magnitude rises by approximately 20%. This pattern aligns with the expectation that a successful breach implies attackers have bypassed existing ZTCs, with servers and user devices emerging as the most probable compromised assets.

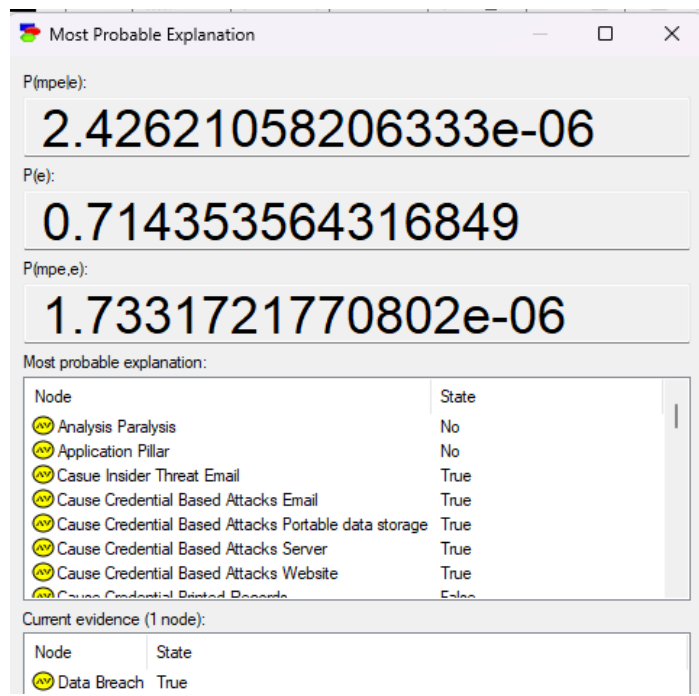

Figure 7 The results of the MPE analysis of scenario 2.

Table 23 The analyzed output of the backward propagation analysis of scenario 2.

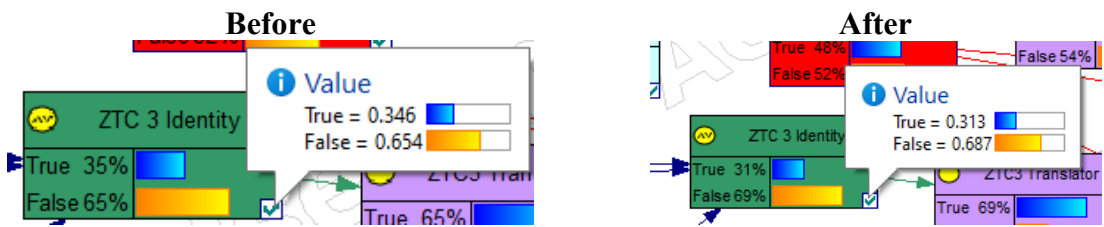

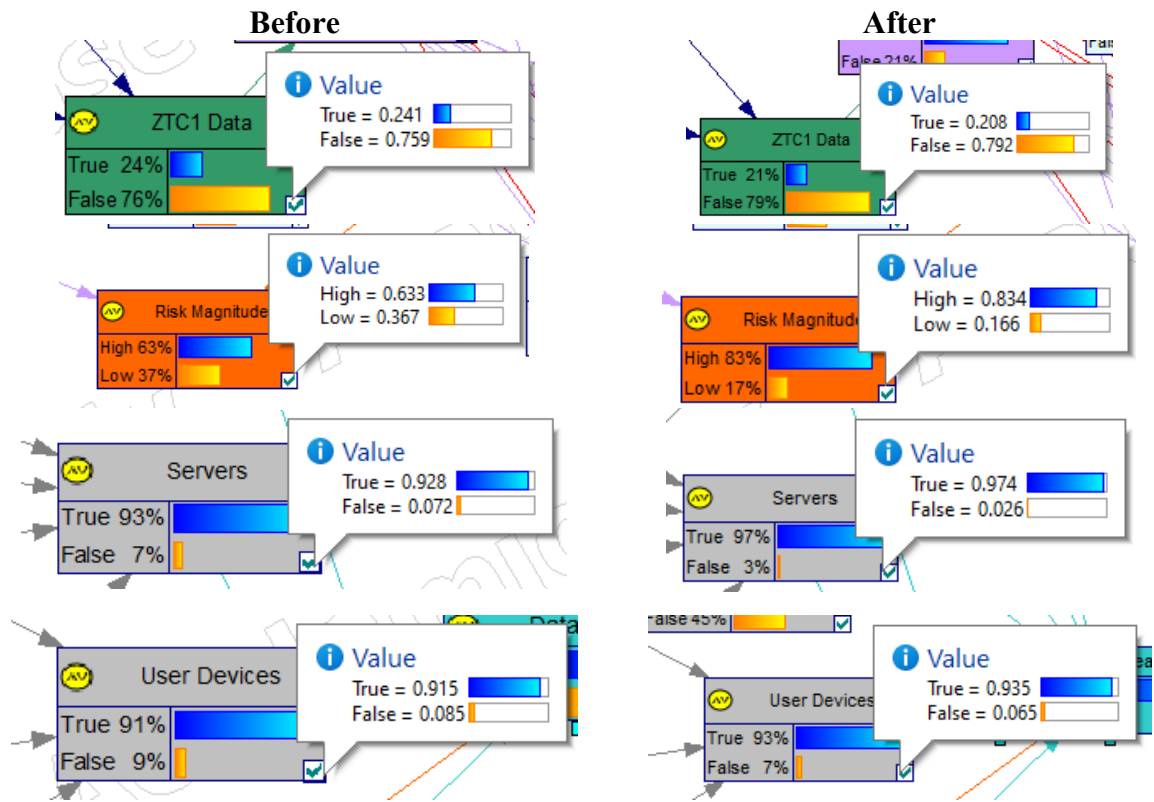

### 4.2.3 Sensitivity Analysis

Sensitivity analysis is used in the Bayesian Network to both validate probability parameters and assess how strongly independent variables influence a given target node under specified conditions. In practice, it enables the construction of tornado graphs that visually rank parent nodes by their impact on a child node, providing actionable insights for SMB managers and cybersecurity practitioners on which factors merit prioritization in real-world decision-making [74, 89]. Within this study, sensitivity analysis is executed in GeNIE for selected key nodes, and the resulting tornado diagrams are examined to interpret variable importance.

### 4.2.3.1 Sensitivity Analysis of Financial Barriers

For the *FinancialBarriers* node, sensitivity analysis is conducted with respect to its parent variables: LimitedBudget, ZTCosts, and *PoorZTBudgetEstimation*. The tornado graphs in Figure 8(a) and 7(b) show how changes in each parent affect the probability that *FinancialBarriers* is either Present or Absent. In these graphs, bar length represents the magnitude of influence of each parent on the target node. Both figures indicate that *ZTCosts* exerts the strongest influence on *FinancialBarriers*, while *PoorZTBudgetEstimation* has the weakest effect, with *LimitedBudget* ranked second. When *ZTCosts* shifts from False (low) to True (high), the probability that *FinancialBarriers* is Present increases from 0.603 to 0.737, whereas *PoorZTBudgetEstimation* only shifts the probability within a narrow range, from 0.611 to 0.618. *LimitedBudget* shows a moderate but meaningful impact between these extremes. These results suggest that using more affordable ZT security measures is the most effective way to reduce financial barriers, followed by increasing the cybersecurity budget in SMBs, while improving budget estimation practices alone has only a marginal effect on mitigating financial constraints.

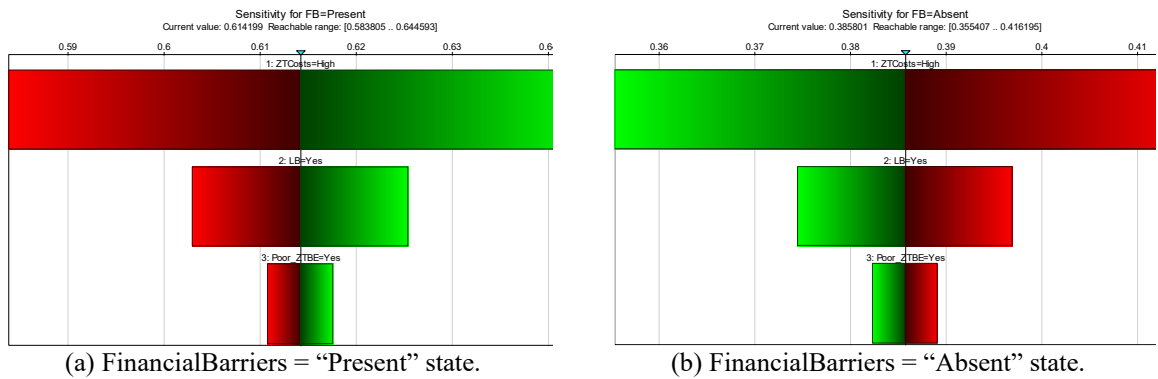

(a) FinancialBarriers = "Present" state.  (b) FinancialBarriers = "Absent" state.

Figure 8 Sensitivity Analysis of FinancialBarriers node.

4.2.3.2 Sensitivity Analysis of Organizational Barriers

The sensitivity analysis for the OrganizationalBarriers node, illustrated in the tornado diagrams in Figure 9(a) and (b), indicates that the inability to hire a qualified security analyst to manage ZT security measures exerts the strongest influence, with an impact range between 0.612 and 0.748. *DifficultLegacy-SystemSubstitution* and *ResistancetoChange* follow as major contributors to organizational complexity, while *LimitedCybersecurityAwareness* and *ZTTechKnowledge* also create organizational hurdles but with comparatively minor effects. The AnalysisParalysis node has been excluded from the sensitivity analysis due to the way the sensitivity analysis algorithm in GeNIE works. The algorithm focuses on nodes and parameters with the maximum derivative impact under the current network configuration, and in certain cases, it distributes the sensitivity impact upstream to grandparent nodes rather than the intermediate node itself. That is why the *LimitedCybersecurityAwareness* and *ZTTechKnowledge* variables are included in the analysis.

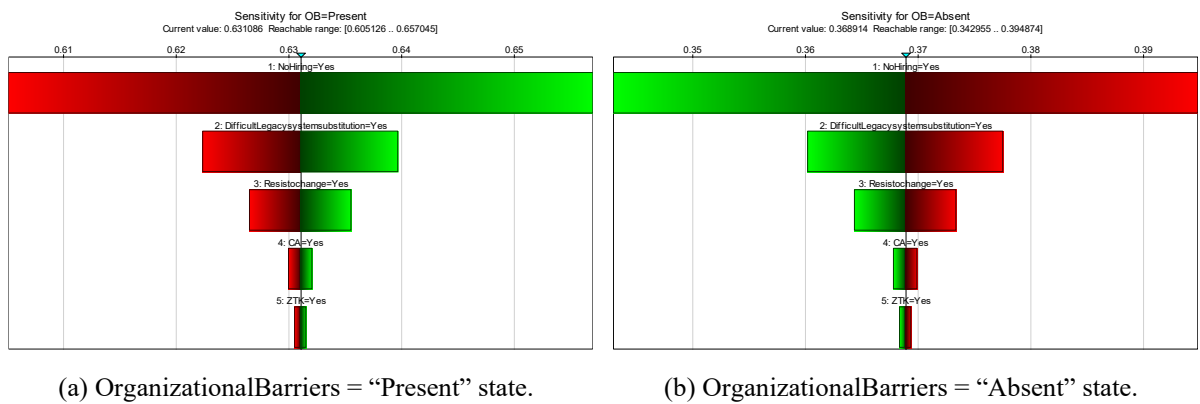

(a) OrganizationalBarriers = "Present" state.  (b) OrganizationalBarriers = "Absent" state.

Figure 9 Sensitivity Analysis of OrganizationalBarriers node.

4.2.3.3 Sensitivity Analysis of ZT Implementation Success

The sensitivity analysis of the *ZTImplementationSuccess* node quantitatively identifies the variables exerting the greatest influence on the success probability of ZT implementation. As illustrated in Figure 10(a), the findings indicate that high ZT costs, shortages of qualified security personnel, and restricted cybersecurity budgets are the primary factors negatively affecting ZT implementation success. Conversely, the ZT security controls that most significantly enhance success include EDR, Device Inventory, RBAC, DE, SSO, MFA, DM, DLP, and WAF, listed in descending order of impact.

These results underscore that financial and organizational limitations represent substantial barriers to achieving successful implementation. Although other elements, such as cyber awareness, WAF, and technical knowledge of ZT, exert comparatively weaker effects in the modeled scenario, their roles remain complementary. Therefore, to maximize the likelihood of successful ZT adoption among SMBs, addressing financial challenges and human resource shortages is essential. Moreover, implementing core ZT controls (e.g., MFA, SSO, DE, RBAC) is critical to ensure operational viability. Finally, cultural readiness and professional training initiatives can meaningfully enhance the overall success of the implementation of ZT.

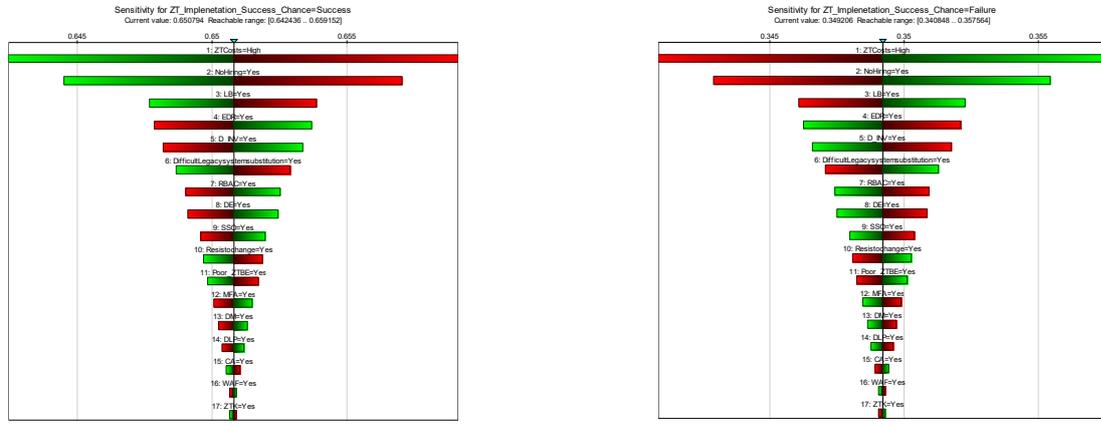

(a) ZTImplementationSuccess = "Success" state.     (b) ZTImplementationSuccess = "Failure" state.

Figure 10 Sensitivity Analysis of ZTImplementationSuccess node.

#### 4.2.3.4 Sensitivity Analysis of Data Breach

The sensitivity analysis of the *DataBreach* node, illustrated in Figures 11(a) and (b), identifies the primary attack vectors most likely to result in data breach incidents. The most influential factors are credential-based attacks, insider threats, physically lost or stolen assets, and phishing, respectively. Among these, phishing demonstrates the weakest effect as a probable cause of data breaches, whereas credential-based and insider attacks represent the most critical risks.

These findings attribute such vulnerabilities primarily to the absence of data encryption, weak access controls, and ineffective identity management practices. To address these root causes, the analysis emphasizes the necessity of implementing ZT security measures. Specifically, adopting data encryption, strengthening RBAC configurations, and enhancing identity and credential mechanisms (e.g., SSO and MFA) are essential steps. In addition, integrating DLP, device management, and EDR solutions provides comprehensive protection against breach-related threats. Collectively, these security measures establish a multilayered defense that substantially reduces the likelihood of data breach incidents within SMB environments.

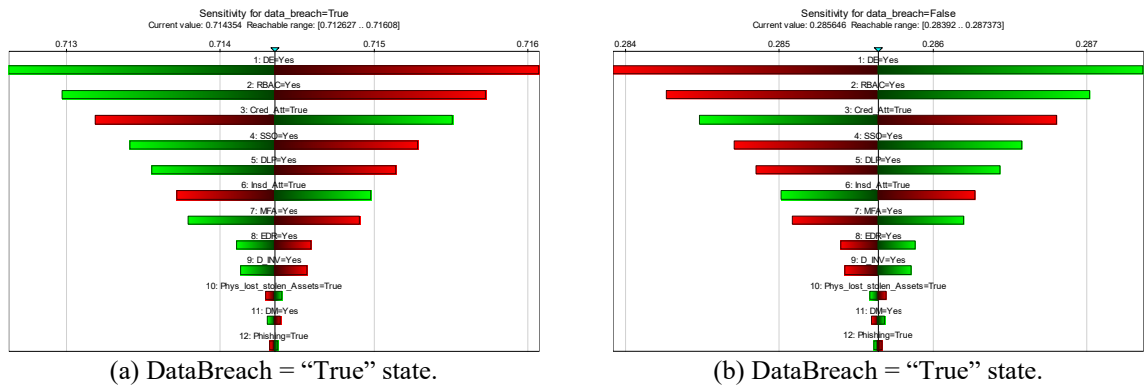

| (a) DataBreach = "True" state. | (b) DataBreach = "True" state. |

Figure 11 Sensitivity Analysis of the DataBreach node.

4.2.3.5 Sensitivity Analysis of Risk Level

The tornado graph of the sensitivity analysis, presented in Figures 12(a) and (b), demonstrates that the overall cybersecurity risk level for SMBs is largely determined by the successful implementation of specific ZT measures. According to the modeled scenario, DE and RBAC exert the greatest positive influence on reducing risk. Additionally, the deployment of robust SSO, effective DLP, and EDR solutions further contributes to lowering SMBs' overall risk exposure.

Conversely, financial and organizational challenges act as amplifiers of cybersecurity risk. High ZT implementation costs, combined with limited cybersecurity budgets, can pose substantial financial obstacles for SMBs. Furthermore, the shortage of qualified security professionals undermines the ability to maintain and operate ZT systems effectively, leaving critical infrastructures vulnerable to cyberattacks. Resistance to change and the complexities of replacing existing IT systems with ZT architecture also represent significant organizational barriers that can heighten risk levels.

In relation to attack vectors, credential-based attacks are identified as the most critical threat influencing the risk node, followed by insider threats. Meanwhile, physically lost or stolen assets and phishing incidents exhibit a lesser but still notable impact on overall risk.

Collectively, these insights highlight the importance for SMBs to prioritize investments in identity management, data encryption, and data loss prevention mechanisms. At the same time, addressing the underlying organizational and economic constraints remains essential for achieving practical and sustainable Zero Trust adoption.

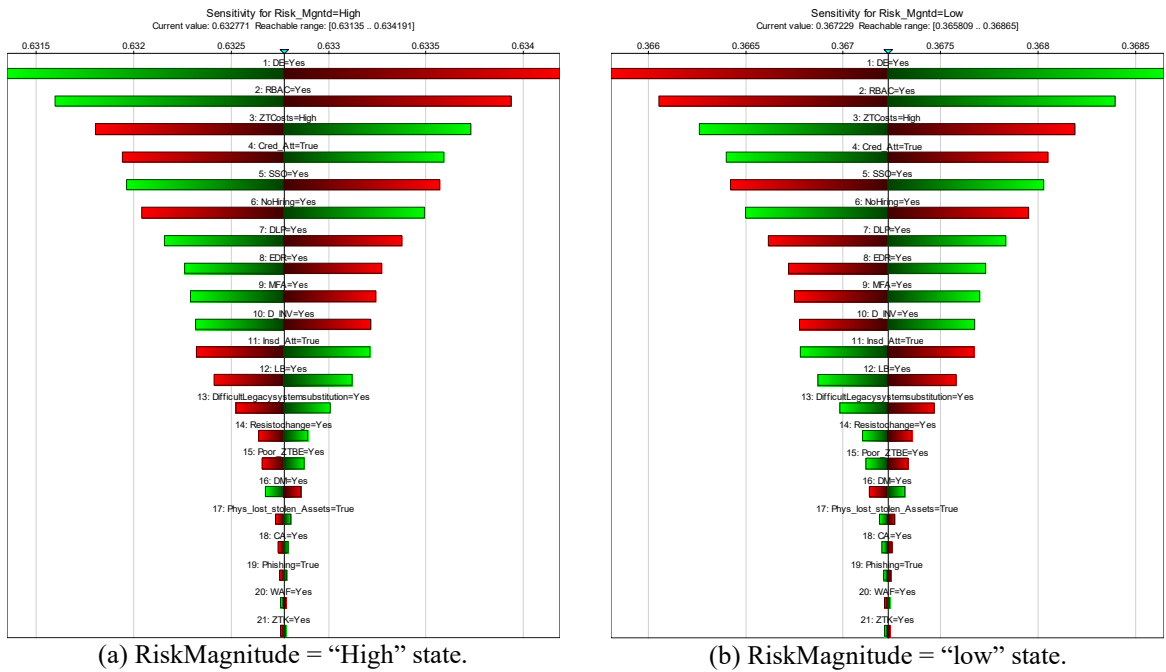

(a) RiskMagnitude = "High" state.  (b) RiskMagnitude = "low" state.

Figure 12 Sensitivity analysis of the response variable RiskMagnitude.

## 5. Research Key Takeaways

The primary objective of this research is to construct a predictive Bayesian Network (BN) model that facilitates scenario-based testing for decision-making in ZTA adoption within SMBs. The model aims to assess both the effectiveness of ZT security measures in safeguarding SMB digital ecosystems against cyberattacks and the influence of organizational and economic constraints on ZTA adoption outcomes. It accounts for inherent business limitations, such as cyber risk management capacity and implementation challenges, and quantifies their impact on the probability of successful ZTA deployment. Simulated attack scenarios enable the estimation of how ZTCs mitigate attack likelihoods, asset compromise probabilities, and data breach incidents, thereby quantifying uncertainty in risk reduction and the maturity of ZT implementation. Through simulation-based analyses, including forward and backward inference, sensitivity analysis, and MPE analysis, the study responded to the research questions outlined in Section 1.

RQ1 investigates the impact of financial and organizational barriers on ZTA adoption success. Results from forward and sensitivity analyses (Table 19; Figures 8–9) reveal that financial and organizational factors substantially hinder reliable ZTA implementation. Elevated costs, limited cybersecurity budgets, lack of qualified personnel, employee resistance to change, and the difficulty of replacing legacy systems were identified as core impediments. Addressing these factors is essential to achieving sustainable implementation performance within SMB environments.

RQ2 explores the Zero Trust pillars and corresponding security measures that enhance SMB cybersecurity. A comprehensive review of academic and industry sources identified four key ZT pillars: Identity, Device, Data, and Application, and nine major security measures: MFA, SSO, RBAC, Device Inventory, Device Management, EDR, DE, DLP, and WAF. The analysis determined that these measures provide adequate maturity levels aligned with SMB budgets and complexity constraints, thereby supporting effective adoption and improved cyber resilience.

RQ3 assesses the effectiveness of ZT security tools in lowering overall risk levels. Forward analysis (Table 20) and sensitivity results (Figure 10) confirm that implementing these tools reduces SMBs' exposure to modeled attack vectors. The tornado graphs further highlight the relative importance of specific ZT tools in minimizing risk magnitude across the ecosystem.

RQ4 examines how individual ZT controls influence the likelihood of data breach incidents. Forward, backward, and MPE analyses demonstrate that ZTC adoption progressively reduces both breach probabilities and total risk. ZTC1 decreases risk to 56%, mainly for servers and websites; ZTC2 lowers it to 55%, improving protection for emails and user devices; and ZTC3 ensures a substantial decline to 40%, extending coverage to portable data storage. Implementation of MFA, SSO, and backup controls through ZTC2 and ZTC3 reduces phishing-related breach probability from 45% to 6%. Although printed records remain vulnerable to insider misuse due to physical access limitations, credential-based, insider, and lost/stolen asset attacks are significantly mitigated under simulated ZTC conditions (Table 21). Backward analysis (Table 22) confirms that strengthening ZTC maturity and implementation success is crucial to maintaining continuous ZTA operability, with servers and user devices prioritized due to their higher compromise potential. The MPE results emphasize that multiple concurrent weaknesses, reflective of a "Swiss cheese" effect, precipitate breaches, illustrating that financial and organizational barriers exacerbate the cascading risks of partial ZT adoption.

RQ5 identifies the most probable attack vectors driving data breach incidents and overall cyber risk. Sensitivity analysis results indicate that credential-based attacks and insider threats are the dominant risk sources, followed by lost or stolen assets and phishing, which exert lesser influence. Consequently, adopting targeted ZT measures, particularly DE, RBAC, MFA, SSO, and DLP, is essential to counter these prevalent vectors and lower SMB cyber risk comprehensively.

In summary, the BN model effectively demonstrates how strategic implementation of ZT controls, coupled with mitigation of financial and organizational barriers, can enhance ZTA success and reduce cyber risk across SMB digital ecosystems.

## 6. Practical Implications and Insights for Stakeholders

The proposed model establishes a foundational framework for implementing ZTA in SMB ecosystems. It identifies and tests appropriate ZT security measures that align with SMBs' operational needs and constraints, demonstrating their effectiveness in defending against malicious cyber threats. Beyond validating technical efficiency, the model accounts for financial and organizational barriers that may impede ZTA implementation, showing its practicality for resource-limited SMBs.

This research encourages SMBs to adopt a proactive cybersecurity mindset by embedding ZTA principles into their risk management strategies. As a holistic and preventive framework, ZTA minimizes dependence on reactive cybersecurity methods found in conventional approaches. Although cyber insurance may help mitigate post-breach financial losses, it remains reactive, offering limited protection against operational disruptions, reputational harm, or compliance violations. Conversely, the proposed model highlights that implementing cost-efficient ZT security tools can serve as a more sustainable, cost-effective alternative, optimizing security investment while reducing long-term exposure to breaches.

The model also enables data-driven decision-making in deploying ZT controls and simplifies compliance with regulatory frameworks such as FISMA, HIPAA, and CMMC. Adopting the selected ZT measures supports SMBs in meeting industry-mandated security requirements, particularly for organizations serving the Department of Defense (DoD) under CMMC guidelines.

From a governance, risk, and compliance (GRC) perspective, the model provides structured guidance for building proactive risk management strategies. Governance ensures consistent enforcement of organizational security policies, while risk management focuses on identifying and prioritizing cyber threats within business contexts. Compliance serves as the regulatory backbone, supported by ZTA's capabilities for continuous verification, stringent access controls, and real-time monitoring, which collectively reduce noncompliance risks and human error. Studies such as Kudrati and Pillai [42] reinforce this alignment, noting that both compliance and ZTA share a common goal, risk minimization.

ZTA simplifies adherence to evolving regulatory requirements through automated policy enforcement and continuous validation of user and device access. For SMBs, which often lack dedicated compliance personnel and cybersecurity awareness, this automation improves compliance outcomes while lowering operational costs. Moreover, governance efforts are streamlined through consistent policy enforcement at each access point, maintaining regulatory alignment and minimizing oversight burdens.

From a risk management perspective, ZTA enhances all four stages of the process, including risk identification, assessment, response, and monitoring.

- Risk Identification: ZTA facilitates the recognition and prioritization of assets, data, and network components, ensuring security objectives are met across all pillars.

- Risk Assessment: Continuous verification via policy decision and enforcement points (PDP and PEP) enables real-time detection of anomalies. Features like MFA, device management, segmentation, and data encryption collectively mitigate unauthorized access and data loss risks.

- Risk Response: ZTA integrates assessment feedback directly into access control decisions, allowing immediate actions, such as granting, denying, or escalating authentication, based on real-time risk evaluations.

- Risk Monitoring and Reporting: Continuous network visibility, event logging, and incident reporting enhance transparency and leadership oversight. Collected security data enriches organizational awareness and informs iterative improvement of cybersecurity strategies.

Overall, the proposed model positions ZTA as both a strategic enabler and a practical risk management framework for SMBs, helping them proactively manage cyber threats, streamline governance and compliance, and strengthen long-term digital resilience.

## 7. Research Limitations, Conclusions, and Future Work

This research faced several inherent limitations arising from the novelty of the ZTA paradigm and data scarcity within the SMB context. Due to the absence of comprehensive datasets and limited empirical evidence, weakly informative priors were employed to estimate posterior probability distributions for selected nodes and to execute Monte Carlo simulations. These priors were derived from the researcher's domain expertise and supplemented with data from cybersecurity forums, aftermarket reports, and industry databases. Since most organizations, particularly SMBs, are still in the early stages of ZTA adoption with modest maturity levels, data availability and quality remain ongoing challenges.

The proprietary dataset used in this study contained over 137,000 cyberattack instances from multiple countries and company sizes. To align with the study's focus, a filtering algorithm was applied to extract 1,486 cases specifically involving U.S.-based SMBs. Although this subset provides a meaningful representation, its moderate size may constrain the statistical reliability of the priors, especially for rare attack categories. While the Advisen dataset offers extensive coverage, its use introduces potential sampling and reporting biases, stemming from update delays and uneven data representation across industries or regions. Moreover, as a proprietary source, it limits replicability, though the proposed model remains adaptable to publicly available datasets.

This study's geographical focus on U.S.-based SMBs stems from definitional inconsistencies across countries and higher data availability in the U.S. context. Nonetheless, the proposed model can be generalized globally by integrating new datasets as more cyber incidents targeting SMBs become documented. The scarcity of large, high-quality datasets in SMB cybersecurity justifies the methodological design adopted here, aligning with constraints typical in this research domain. Additionally, the developed model is static, meaning it does not capture temporal dependencies among ZTCs or variables, an aspect that could be enhanced in future work. Given that ZTA adoption is a gradual, multi-phase process, the present research centers on financial and organizational risk factors that most strongly influence implementation success.

Despite these limitations, this study contributes a comprehensive, evidence-based model for evaluating ZTA implementation success and its effect on SMB cybersecurity resilience. SMBs form the backbone of developed economies, yet their limited resources, technical capacity, and cybersecurity awareness make them highly susceptible to attacks. The proposed Bayesian model provides a structured mechanism to simulate decision-making scenarios, test the influence of business and technical factors, and identify ZT measures most relevant to SMBs' operational and financial constraints. By enabling quantitative testing of "what-if" conditions, the model supports informed decisions in adopting ZT controls while highlighting strategies to mitigate financial and organizational barriers.

Empirical analyses revealed that affordable ZT measures, such as MFA, SSO, RBAC, DE, and DLP, significantly strengthen SMBs' ability to withstand cyber threats and reduce the overall risk level. The results underscore that ZTA, when implemented progressively in alignment with business constraints, enhances cyber resilience by enabling proactive defense mechanisms and rapid response to cyber incidents. Notably, the study represents one of the first attempts to predict the likelihood of successful ZTA adoption in SMBs and to quantify the effect of ZT controls on overall risk levels.

Building on these findings, several avenues for future research are recommended. Expanding the model to incorporate additional risk dimensions, such as technical, supply chain, and operational dependencies, would enhance its comprehensiveness. Introducing temporal dynamics to form a Dynamic Bayesian Network (DBN) could better represent evolving cyber threats and interdependencies among variables. Feedback loops could enable adaptive learning, improving model relevance and accuracy in real-time environments. Moreover, knowledge elicitation from subject matter experts can refine prior estimations and offset data limitations, while integrating external threat intelligence feeds can continuously align the model with the current cyber landscape.

Further research should also quantify the economic implications of ZTA adoption, including the estimation of implementation costs and the calculation of Return On Investment (ROI) for specific ZT tools. Performing cost-benefit analyses would help identify high-impact controls and demonstrate the financial feasibility of ZTA for SMBs. Finally, future modeling efforts could explore dependencies among attack vectors, reflecting real-world attack chains where adversaries exploit multiple pathways simultaneously.

In conclusion, the study establishes a foundational framework for predicting ZTA adoption success and quantifying its risk mitigation potential in SMBs. It advances cybersecurity modeling by combining organizational, financial, and technical perspectives within a probabilistic structure. The proposed model offers both academic and practical value, serving as a decision-support tool for SMBs seeking to strengthen cyber defenses and for researchers aiming to extend the frontiers of Zero Trust-based risk management.

# Acknowledgements

The authors gratefully acknowledge the Insurance and Financial Services Center at Old Dominion University's Strome College of Business for funding the acquisition of the Advisen cyber loss dataset used in this research. The authors also acknowledge the use of the GeNIE Modeler$^©$ software, developed and maintained by BayesFusion, LLC, which is available for academic use at http://www.bayesfusion.com/.

# Author contributions: CRediT

Conceptualization: AA, Data curation: AA and MM, Formal analysis: AA, Investigation: AA and BE, Methodology: AA, Resources: BE and MM, Supervision: BE, Validation: AA and BE, Visualization: AA and BE, Writing – original draft: AA, Writing – review and editing: AA, BE and MM.

# Funding sources

This research did not receive any specific grant from funding agencies in the public, commercial, or not-for-profit sectors.

# Appendices

**Appendix A**

Prior predictive checks are used to validate whether the assigned weakly informative Beta priors yield plausible behavior before incorporating data. Following the approach discussed by Gabry, Simpson [149], data are simulated from the prior predictive distribution to ensure that the induced pillar activation probabilities are reasonable (e.g., not concentrated near 0 or 1 under weakly informative priors). Table A.1 shows the implied prior predictive probabilities for each ZT pillar. The Identity and Device pillars have median activation probabilities of 0.58 and 0.54, respectively, with 95% credible intervals of [0.43–0.72] and [0.43–0.65], indicating that, a priori, these pillars are more likely than not to be active and provide meaningful protection. By contrast, the Application pillar yields a median of 0.11, with a plausible range from 0.05 to 0.19, signaling a more modest expected contribution under current SMB adoption patterns.

Table A.1 Prior predictive summary of ZT Pillars nodes.

| Variable Name | Mean | Median | SD | $q_{2.5\%}$ | $q_{97.5\%}$ |
|---|---|---|---|---|---|
| Data | 0.293243 | 0.290712 | 0.053918 | 0.194157 | 0.405239 |
| Identity | 0.580737 | 0.582392 | 0.073481 | 0.432644 | 0.72104 |
| Device | 0.540507 | 0.541408 | 0.056392 | 0.427927 | 0.650311 |
| Application | 0.111998 | 0.108074 | 0.035985 | 0.053059 | 0.193741 |

**Algorithm 1: SMBs Names Text Search**

```
1: Begin
2:    Load SMB file (List-of-SME-Companies.csv)
3:    Load Advisen incidents file (AdivisenData.csv)
4:    STANDARDIZE column names in both datasets
5:    Check that SMB file has ["COMPANY NAME", "TOTAL EMPLOYEES", "HQ COUNTRY"]
6:    Check that Incidents file has ["COMPANY_NAME", "COUNTRY_CODE"]
7:    for each SMB record:
8:        Normalize HQ Country using function normalize_country
9:        if HQ Country == "United States":
10:           Keep record
11:       else:
12:           Discard record
13:   Apply clean_company_name to SMB company names
14:       STORE all normalized SMB names in a set
15:   Filter Advisen incidents where COUNTRY_CODE == "USA"
16:   Apply clean_company_name to incident company names
17:   for each incident:
18:       if CLEAN_NAME in SMB name set:
19:           Add to matched_incidents
20:   if "MSCAD_ID" exists in matched_incidents:
21:       Create normalized MSCAD_ID (string, trimmed)
22:       Identify duplicates with same MSCAD_ID
23:       Export duplicate rows to CSV for inspection
24:       Remove duplicates, keep first occurrence
25:   Export matched_incidents (without helper columns) to CSV
26:   print summary
27: end
```

Figure A.1 A pseudo-code for Algorithm 1.

**Algorithm 2:** Estimate Prior Probabilities of Cyber Attack Types

```
1: function CALCULATE_PRIORS(dataset df, attack_rules)
2:     STANDARDIZE column names in df by stripping spaces and converting to uppercase
3:     total_instances ← LENGTH(df)
4:     distinct_attacks ← LENGTH(attack_rules)
5:     results ← ∅
6:     for each (attack_type, rules) in attack_rules do
7:        mask ← Boolean array initialized as False for all rows
8:        if attack_type == "Insider Threats" then
9:         mask ← (df["CASE_TYPE"] contains "Insider Threat")
10:          if not mask then
11:            if df["PRODUCT_SERVICE_INVOLVED"] == "Internal" then
12:               mask ← True
13:            else if df["PRODUCT_SERVICE_INVOLVED"] is blank then
14:               mask ← (df["CASE_DESCRIPTION"] contains any keyword in rules["CASE_DESCRIPTION"])
15:            end if
16:          end if
17:        else       ▷ For all other attack types
18:           if rules["CASE_TYPE"] ≠ ∅ then
19:               mask ← mask OR (df["CASE_TYPE"] contains any keyword in rules["CASE_TYPE"])
20:           end if
21:            if rules["CASE_DESCRIPTION"] ≠ ∅ then
22:                mask ← mask OR (df["CASE_DESCRIPTION"] contains any keyword in rules["CASE_DESCRIPTION"])
23:            end if
24:            if rules["PRODUCT_SERVICE_INVOLVED"] ≠ ∅ then
25:                mask ← mask AND (df["PRODUCT_SERVICE_INVOLVED"] in rules["PRODUCT_SERVICE_INVOLVED"])
26:            end if
27:          end if
28:        filtered_df ← df[mask]
29:        attack_count ← LENGTH(filtered_df)
30:         prior_prob ← (attack_count + 1) / (total_instances + distinct_attacks)   ▷ Laplace smoothing
31:        APPEND(attack_type, prior_prob, attack_count) to results
32:     end for
33:     return results
34: end function
```

Figure A.2 A pseudo-code for Algorithm 2.

**Algorithm 3:** Estimate Conditional Probabilities of Assets Given Attack Types

```
1: function CALCULATE_CONDITIONALS(attack_files, asset_mapping)
2:     results ← ∅
3:     for attack in attack_files do
4:         df ← READ_CSV(full_assignment_table.file)
5:         if "PROXIMATE_CAUSE" ∉ df.columns then
6:             RAISE Error("Column missing")
7:         end if
8:         df["ASSET"] ← MAP(df["PROXIMATE_CAUSE"], asset_mapping)
9:         df ← DROP_ROWS_WHERE(df["ASSET"] = NULL)
10:        total ← LENGTH(df)
11:        counts ← VALUE_COUNTS(df["ASSET"])
12:        for asset in counts.keys() do
13:            prob ← counts[asset] / total
14:            APPEND(results, (attack, asset, counts[asset], prob))
15:        end for
16:        SAVE(results for attack) to CSV
17:    end for
18: combined ← CONCAT(results across attacks)
19: SAVE(combined) to "combined_cpt.csv"
20: return combined
21: end function
```

Figure A.3 A pseudo-code for Algorithm 3.

Table A.2 The assigned Beta priors between Attack nodes, ZTC nodes, and intermediate nodes.

| Attack Name | Asset | ZTC Name | Assigned Informative Beta Distribution | Mean | Rationale and Reference |
|---|---|---|---|---|---|
| Phishing | Server | ZTC1 | (7,3) | 0.70 | Data controls (e.g., DE, DLP) can inhibit 70% of phishing attempts on servers |
| Phishing | Server | ZTC3 | (6,4) | 0.60 | Identity controls (e.g., MFA, RBAC) can inhibit 60% of phishing on servers |
| Phishing | Email | ZTC2 | (7,3) | 0.70 | Device controls (EDR, DM, etc.) can block 70% of endpoint phishing |
| Phishing | Email | ZTC3 | (8,2) | 0.80 | Identity controls can effectively block up to 80% of phishing attacks |
| Insider Threat | Server | ZTC1 | (5,5) | 0.50 | Data controls are moderately effective for insider threats |
| Insider Threat | Server | ZTC3 | (6,4) | 0.60 | Identity controls about 60% of insider threats |
| Insider Threat | Email | ZTC2 | (5,5) | 0.50 | Device controls are moderately effective for insider threats |

| Attack Name | Asset | ZTC Name | Assigned Informative Beta Distribution | Mean | Rationale and Reference |
|---|---|---|---|---|---|
| Insider Threat | Email | ZTC3 | (6,4) | 0.60 | Identity controls can help restrict unauthorized access, especially for insider attackers |
| Insider Threat | Website | ZTC1 | (5,5) | 0.50 | Data controls on web servers have a moderate effect. It is mostly effective on data manipulation/ exfiltration. |
| Insider Threat | Website | ZTC3 | (6,4) | 0.60 | Identity controls limit credential abuse on apps/ websites. |
| Insider Threat | Printed Records | ZTC3 | (3,7) | 0.30 | Physical document controls by IAM are limited and could be more beneficial for digital assets |
| Insider Threat | User Devices | ZTC1 | (5,5) | 0.50 | Data controls have a moderate mitigating effect on endpoint data leakage. |
| Insider Threat | User Devices | ZTC2 | (6,4) | 0.70 | Device controls can reduce up to 70% of endpoint insider threats |
| Insider Threat | User Devices | ZTC3 | (6,4) | 0.60 | Identity measures restrict lateral movement and privilege abuse |
| Insider Threat | Software | ZTC3 | (6,4) | 0.60 | IAM reduces privilege misuse in the application layer. |
| Insider Threat | Software | ZTC4 | (4,6) | 0.40 | Application-level security is modest for insider threats. |
| Insider Threat | Portable Data Storage Instruments | ZTC1 | (4,6) | 0.40 | ZT Data measures somewhat effectively for protecting data stored on USBs. |
| Physically Lost or Stolen | Server | ZTC1 | (4,6) | 0.40 | ZT Data controls moderately protect servers. |
| Physically Lost or Stolen | Server | ZTC3 | (8,2) | 0.80 | ZT Identity measures are effective in protecting servers |
| Physically Lost or Stolen | Printed Records | ZTC3 | (3,7) | 0.30 | Identity controls have a low effect (30%), as they depend on other physical controls implemented by organizations |
| Physically Lost or Stolen | User Devices | ZTC1 | (5,5) | 0.50 | Encryption is needed for device loss; thus, moderate protection is assumed |
| Physically Lost or Stolen | User Devices | ZTC2 | (8,2) | 0.80 | Device controls can block up to 80% loss/ theft of assets |
| Physically Lost or Stolen | User Devices | ZTC3 | (4,6) | 0.40 | IAM helps with lost devices through kill switches. |
| Physically Lost or Stolen | Portable Data Storage Instruments | ZTC1 | (4,6) | 0.40 | DE on portable storage tools is modestly effective |
| Credential-Based Attacks | Server | ZTC1 | (5,5) | 0.50 | Data measures have moderate for credential attacks |
| Credential-Based Attacks | Server | ZTC3 | (7,3) | 0.70 | Identity controls can block up to 80% of credential-based attacks on servers |

| Attack Name | Asset | ZTC Name | Assigned Informative Beta Distribution | Mean | Rationale and Reference |
|---|---|---|---|---|---|
| Credential-Based Attacks | Email | ZTC2 | (7,3) | 0.70 | Device measures are highly effective for endpoints' protection |
| Credential-Based Attacks | Email | ZTC3 | (8,2) | 0.80 | Identity controls can block up to 80% of credential-based attacks on emails |
| Credential-Based Attacks | Website | ZTC1 | (5,5) | 0.50 | Data controls on the website have a moderate impact |
| Credential-Based Attacks | Website | ZTC3 | (7,3) | 0.70 | Identity controls (e.g., MFA, SSO) on apps block a large share of credential compromise |
| Credential-Based Attacks | User Devices | ZTC1 | (5,5) | 0.50 | DE and DLP have a moderate effect on credential misuse on devices |
| Credential-Based Attacks | User Devices | ZTC2 | (7,3) | 0.70 | Device controls can block up to 70% of credential attacks on endpoints |
| Credential-Based Attacks | User Devices | ZTC3 | (8,2) | 0.80 | The Identity pillar is highly effective for device login |
| Credential-Based Attacks | Portable Data Storage Instruments | ZTC1 | (4,6) | 0.40 | DE on media is only a partial defense for stolen credentials |
| Credential-Based Attacks | Printed Records | ZTC3 | (16,4) | 0.80 | ZT Identity controls are effective in mitigating credential-based attacks |

Table A.3 The MC summary of the causal link strengths between the phishing attack node and intermediate nodes.

| Attack | Asset | ZTC | Mean | Median | SD | $q_{2.5\%}$ | $q_{97.5\%}$ |
|---|---|---|---|---|---|---|---|
| Phishing | Server | ZTC1 | 0.3124 | 0.3116 | 0.0727 | 0.1753 | 0.4571 |
| Phishing | Server | ZTC3 | 0.4707 | 0.4717 | 0.1156 | 0.2456 | 0.6924 |
| Phishing | Email | ZTC2 | 0.5291 | 0.5359 | 0.1034 | 0.3127 | 0.7093 |
| Phishing | Email | ZTC3 | 0.6165 | 0.6233 | 0.1051 | 0.3950 | 0.8015 |

Table A.4 The MC summary of the causal link strengths between the insider threat attack node and intermediate nodes.

| Attack | Asset | ZTC | Mean | Median | SD | $q_{2.5\%}$ | $q_{97.5\%}$ |
|---|---|---|---|---|---|---|---|
| Insider Threat | Server | ZTC1 | 0.2337 | 0.2303 | 0.0707 | 0.1076 | 0.3799 |
| Insider Threat | Server | ZTC3 | 0.4718 | 0.4729 | 0.1173 | 0.2429 | 0.6966 |
| Insider Threat | Email | ZTC2 | 0.3880 | 0.3863 | 0.1088 | 0.1811 | 0.6003 |
| Insider Threat | Email | ZTC3 | 0.4724 | 0.4735 | 0.1165 | 0.2440 | 0.6963 |
| Insider Threat | Website | ZTC1 | 0.2352 | 0.2311 | 0.0711 | 0.1077 | 0.3837 |

| Attack | Asset | ZTC | Mean | Median | SD | $q_{2.5\%}$ | $q_{97.5\%}$ |
| --- | --- | --- | --- | --- | --- | --- | --- |
| Insider Threat | Website | ZTC3 | 0.4695 | 0.4697 | 0.1154 | 0.2462 | 0.6895 |
| Insider Threat | Printed Records | ZTC3 | 0.2531 | 0.2427 | 0.1042 | 0.0827 | 0.4834 |
| Insider Threat | User Devices | ZTC1 | 0.2343 | 0.2310 | 0.0706 | 0.1076 | 0.3826 |
| Insider Threat | User Devices | ZTC2 | 0.4566 | 0.4591 | 0.1085 | 0.2406 | 0.6601 |
| Insider Threat | User Devices | ZTC3 | 0.4717 | 0.4738 | 0.1167 | 0.2424 | 0.6934 |
| Insider Threat | Software | ZTC3 | 0.4717 | 0.4718 | 0.1160 | 0.2470 | 0.6952 |
| Insider Threat | Software | ZTC4 | 0.1168 | 0.1117 | 0.0446 | 0.0449 | 0.2164 |
| Insider Threat | Portable Data Storage Instruments | ZTC1 | 0.1956 | 0.1904 | 0.0671 | 0.0792 | 0.3383 |

Table A.5 The MC summary of the causal link strength between the physically lost or stolen attack node and the intermediate node.

| Attack | Asset | ZTC | Mean | Median | SD | $q_{2.5\%}$ | $q_{97.5\%}$ |
| --- | --- | --- | --- | --- | --- | --- | --- |
| Physically Lost or Stolen | Server | ZTC1 | 0.1961 | 0.1905 | 0.0677 | 0.0791 | 0.3384 |
| Physically Lost or Stolen | Server | ZTC3 | 0.6152 | 0.6216 | 0.1045 | 0.3956 | 0.7990 |
| Physically Lost or Stolen | Printed Records | ZTC3 | 0.2557 | 0.2449 | 0.1046 | 0.0840 | 0.4854 |
| Physically Lost or Stolen | User Devices | ZTC1 | 0.2339 | 0.2305 | 0.0704 | 0.1067 | 0.3785 |
| Physically Lost or Stolen | User Devices | ZTC2 | 0.5975 | 0.6064 | 0.0941 | 0.3919 | 0.7554 |
| Physically Lost or Stolen | User Devices | ZTC3 | 0.3270 | 0.3196 | 0.1116 | 0.1312 | 0.5606 |
| Physically Lost or Stolen | Portable Data Storage Instruments | ZTC1 | 0.1957 | 0.1910 | 0.0672 | 0.0808 | 0.3377 |

Table A.6 The MC summary of the causal link strength between the credential-based attack node and intermediate nodes.

| Attack | Asset | ZTC | Mean | Median | SD | $q_{2.5\%}$ | $q_{97.5\%}$ |
|---|---|---|---|---|---|---|---|
| Credential-Based | Server | ZTC1 | 0.2348 | 0.2309 | 0.0707 | 0.1085 | 0.3822 |
| Credential-Based | Server | ZTC3 | 0.5445 | 0.5486 | 0.1132 | 0.3141 | 0.7532 |
| Credential-Based | Email | ZTC2 | 0.5276 | 0.5344 | 0.1033 | 0.3118 | 0.7115 |
| Credential-Based | Email | ZTC3 | 0.6141 | 0.6214 | 0.1057 | 0.3918 | 0.8015 |
| Credential-Based | Website | ZTC1 | 0.2348 | 0.2311 | 0.0705 | 0.1084 | 0.3806 |
| Credential-Based | Website | ZTC3 | 0.5432 | 0.5480 | 0.1137 | 0.3113 | 0.7523 |
| Credential-Based | Printed Records | ZTC3 | 0.6156 | 0.6188 | 0.0873 | 0.4379 | 0.7772 |
| Credential-Based | User Devices | ZTC1 | 0.2338 | 0.2304 | 0.0709 | 0.1068 | 0.3798 |
| Credential-Based | User Devices | ZTC2 | 0.5277 | 0.5340 | 0.1030 | 0.3146 | 0.7078 |
| Credential-Based | User Devices | ZTC3 | 0.6148 | 0.6207 | 0.1055 | 0.3933 | 0.8018 |
| Credential-Based | Portable Data Storage Instruments | ZTC1 | 0.1950 | 0.1900 | 0.0669 | 0.0795 | 0.3387 |

Table A.7 The assigned strength values between intermediate nodes and asset nodes.

| Attack Type | Asset | Assigned Prior Beta Distribution | Rationale |
|---|---|---|---|
| Phishing | Email | (9,1) | A compromise is likely if phishing gets through |
| Insider Threat | Email | (7,3) | Risk exists, but may be limited by monitoring/culture |
| Credential-Based | Email | (9,1) | Credential theft incidents usually grant full access |
| Credential-Based | Server | (9,1) | Servers are highly at risk post-credential theft |
| Insider Threat | Server | (7,3) | moderate likelihood if the insider has privileged access |
| Phishing | Server | (8,2) | Phishing often targets server access |
| Physically Lost or Stolen | Server | (6,4) | Data centers have physical security, but theft is still possible |
| Credential-Based | Website | (8,2) | Websites are exposed after credential attacks |
| Insider Threat | Website | (7,3) | A privileged insider can compromise web servers |
| Credential-Based | Printed Records | (6,4) | Paper records can be targeted, but extra steps are required |
| Insider Threat | Printed Records | (5,5) | It depends heavily on organization controls |
| Physically Lost or Stolen | Printed Records | (7,3) | Loss/theft likely leads to exposure. |

| Insider Threat | Software | (6,4) | Software sabotage by an insider is possible but mitigated by code reviews |
| --- | --- | --- | --- |
| Credential-Based | User Devices | (8,2) | Credential theft leads to endpoint compromise |
| Physically Lost or Stolen | User Devices | (8,2) | A lost/stolen device is a compromise risk |
| Insider Threat | User Devices | (7,3) | An insider can attack personal endpoints |
| Credential-Based | Portable Data Storage Instruments | (8,2) | Credential theft exposes portable storage |
| Physically Lost or Stolen | Portable Data Storage Instruments | (8,2) | Lost/stolen media likely exposes data |
| Insider Threat | Portable Data Storage Instruments | (7,3) | An insider attack can easily access the media. |

Table A.8 The posterior mean and SD are used to fit Beta distributions for parent nodes.

| Attack | Asset | Parent Posterior Mean | Parent Posterior SD | Parent Fitted Alpha | Parent Fitted Beta |
| --- | --- | --- | --- | --- | --- |
| Phishing | Email | 0.5733 | 0.1129 | 10.4338 | 7.7671 |
| Insider Threat | Email | 0.4298 | 0.1206 | 6.8108 | 9.0347 |
| Credential-Based | Email | 0.5717 | 0.1132 | 10.3437 | 7.7507 |
| Credential-Based | Server | 0.3895 | 0.1815 | 2.4230 | 3.7977 |
| Insider Threat | Server | 0.3527 | 0.1534 | 3.0693 | 5.6318 |
| Phishing | Server | 0.3917 | 0.1244 | 5.6361 | 8.7526 |
| Physically Lost or Stolen | Server | 0.4056 | 0.2277 | 1.4807 | 2.1699 |
| Credential-Based | Website | 0.3898 | 0.1813 | 2.4305 | 3.8053 |
| Insider Threat | Website | 0.3525 | 0.1518 | 3.1370 | 5.7626 |
| Credential-Based | Printed Records | 0.6163 | 0.0870 | 18.6180 | 11.5901 |
| Insider Threat | Printed Records | 0.2532 | 0.1041 | 4.1653 | 12.2825 |
| Physically Lost or Stolen | Printed Records | 0.2546 | 0.1044 | 4.1768 | 12.2257 |
| Insider Threat | Software | 0.2940 | 0.1978 | 1.2665 | 3.0412 |
| Credential-Based | User Devices | 0.4584 | 0.1882 | 2.7564 | 3.2565 |

| Attack | Asset | Parent Posterior Mean | Parent Posterior SD | Parent Fitted Alpha | Parent Fitted Beta |
|---|---|---|---|---|---|
| Physically Lost or Stolen | User Devices | 0.3867 | 0.1802 | 2.4374 | 3.8658 |
| Insider Threat | User Devices | 0.3876 | 0.1478 | 3.8237 | 6.0406 |
| Credential-Based | Portable Data Storage Instruments | 0.1949 | 0.0670 | 6.6227 | 27.3593 |
| Physically Lost or Stolen | Portable Data Storage Instruments | 0.1960 | 0.0674 | 6.6080 | 27.1089 |
| Insider Threat | Portable Data Storage Instruments | 0.1955 | 0.0667 | 6.7207 | 27.6614 |